\documentclass[10pt,journal,compsoc]{IEEEtran}
\usepackage{mdwlist}

\usepackage{booktabs} 
\usepackage{diagbox}
\usepackage{balance}

\usepackage{colortbl}
\usepackage{amsmath}

\usepackage{enumitem}
\usepackage{fancyvrb}
\usepackage{epstopdf}
\usepackage{color}
\usepackage{amsfonts}
\usepackage{listings}
\usepackage{graphicx}
\usepackage{algorithm}
\definecolor{darkblue}{rgb}{0.0, 0.0, 0.55}
\usepackage{verbatim}
\usepackage{multirow}
\usepackage{framed}
\usepackage{caption}
\DeclareCaptionType{copyrightbox}
\usepackage{subcaption}
\usepackage{pifont}
\usepackage{soul}
\usepackage{mdframed}

\usepackage{hyperref}

\usepackage[font={small,bf}]{caption}

\usepackage{multicol}

\usepackage{algpseudocode}
\usepackage{fancyvrb}
\usepackage{soul}
\algtext*{EndWhile}
\algtext*{EndIf}

\newcommand{\commentty}[1]{{\color{blue} \sf (TY: #1)}}

\newcommand{\subB}{\vspace{3pt}\noindent\textbf}

\newcommand{\ignore}[1]{}
\definecolor{Light}{gray}{.85}

\usepackage{xcolor}
\usepackage{newverbs}

\newverbcommand{\cverb}{\color{red}}{}
\newverbcommand{\bverb}
  {\begin{lrbox}{\verbbox}}
  {\end{lrbox}\colorbox{gray!30}{\box\verbbox}}

\newcommand{\Name}{DinoDroid}

\newcommand{\yuc}[1]{#1}

\usepackage{url}

\begin{document}
%
\title{
{\Name{}: Testing Android Apps Using Deep Q-Networks}
}
%
%
%
%

\author{Yu~Zhao,~\IEEEmembership{Member,~IEEE,}
        Brent~Harrison,~\IEEEmembership{Member,~IEEE,}
        and~Tingting~Yu,~\IEEEmembership{Member,~IEEE}
\IEEEcompsocitemizethanks{\IEEEcompsocthanksitem Yu Zhao was with the Department
of Computer Science and Computer Science and Cybersecurity, University of Central Missouri, Warrensburg,
MO, 64093.\protect\\
E-mail: yzhao@ucmo.edu
\IEEEcompsocthanksitem Brent Harrison was with the Department
of Computer Science, University of Kentucky, Lexington,
KY, 40506.\protect\\
E-mail: harrison@cs.uky.edu.
\IEEEcompsocthanksitem Tingting Yu (corresponding author) was with the Department
of Electrical Engineering and Computer Science, University of Cincinnati, Cincinnati,
OH, 45221.\protect\\
E-mail: yutt@ucmail.uc.edu}
}

\IEEEtitleabstractindextext{%
\begin{abstract}
The large demand of mobile devices creates 
significant concerns about the quality of mobile applications (apps). 
Developers need to guarantee the quality of mobile apps before it is
released to the market. There have been many approaches using 
different strategies to test the GUI of mobile apps. 
However,  they still need improvement due to their limited effectiveness. 
In this paper, we propose \Name{}, an approach based on 
deep Q-networks to automate testing of Android apps.  
\Name{} learns a behavior model from a set of existing apps and 
the learned model can be used to explore and generate tests for new apps.
\Name{} is able to capture the fine-grained details of GUI events 
(e.g., the content of GUI widgets) and use
them as features that are fed into deep neural network, which
acts as the agent to guide app exploration. 
\Name{} automatically adapts the learned model during the
exploration without the need of any modeling strategies or pre-defined rules.
We conduct experiments on 64 open-source Android apps. 
The results showed that \Name{} outperforms 
existing Android testing tools in terms of code coverage and bug detection. 
\end{abstract}

}

\maketitle

\section{Introduction}
\label{sec:introduction}

Mobile applications (apps) have become extremely popular with about
three million apps in Google Play's app
store~\cite{googleplaylink}.  
The increase in app complexity has created significant 
concerns about the quality of apps.  Also,  because of 
the rapid releasing cycle of apps and limited human resources,
it is difficult for developers  to manually construct test cases. 
Therefore, different automated mobile app testing techniques have been 
developed and applied ~\cite{choudhary2015automated}.

Test cases for mobile apps are often represented by
sequences of GUI events \footnote{In our setting, an event
refers to an executable GUI widget associated with
an action type (e.g., click, scroll, edit, swipe, etc).}
to mimic the interactions between users and apps. 
The goal of an automated test generator is generating such event
sequences to achieve high code
coverage and/or detecting bugs. A successful test generator
is able to exercise the correct GUI widget on the
current app page, so that when exercising that widget, it can 
bring the app to a new page, leading to the exploration of new 
events. However, existing mobile app
testing tools often explore a limited set of events because they have
limited capability of understanding which GUI events would expand
the exploration like humans do.  This can lead to automated test 
generators performing unnecessary actions that are unlikely to lead to new 
coverage or detect new bugs.

Many automated GUI testing approaches for mobile apps have been proposed,
such as random testing~\cite{monkey, machiry2013dynodroid} and model-based testing~\cite{amalfitano2012using, amalfitano2015mobiguitar, azim2013targeted}. 
Random testing (e.g., Monkey~\cite{monkey})
is popular in testing mobile apps because
of its simplicity and availability.  It generates 
tests by sending thousands of GUI events 
per second to the app. While random testing 
can sometimes be effective, it is difficult to explore
hard-to-reach events to drive the app to new pages because of 
the natural of randomness. 
Model-based testing~\cite{azim2013targeted,su2017guided} can improve code coverage by employing 
pre-defined strategies or rules to guide the app exploration.
For example, 
A3E~\cite{azim2013targeted}  employs depth-first search (DFS) 
to explore the model of an app under test (AUT) based on event-flow across
app pages. 
Stoat~\cite{su2017guided} utilizes a stochastic Finite State Machine model to
describe the behavior of  AUT and then utilizes Markov Chain Monte
Carlo sampling~\cite{chib1995understanding} to guide the testing. 
%
%
%
However, model-based testing often relies on human-designed models 
and it is almost impossible to precisely model an app's behavior. 
Also, many techniques apply pre-defined rules to the model for improving testing.
For example, Stoat~\cite{su2017guided} designed rules to assign each event an execution weight
in order to speed up exploration. However,   these pre-defined rules are often derived from limited observations and may not generalize to a wide categories of apps.


To summarize, the inherent limitation of the above techniques is that they do not \emph{automatically}
understand  GUI layout and the content of the GUI elements, so it is difficult for them to exercise
the most effective events that will bring the app into new states.  
Recently, machine learning techniques have been proposed to perform GUI testing
in mobile apps~\cite{koroglu2018qbe, pan2020reinforcement, li2019humanoid, degott2019learning}. For example,  Humanoid~\cite{li2019humanoid} uses
deep learning to learn from human-generated interaction traces and uses the learned model to guide test 
generation as a human tester. However, this approach relies on human-generated
datasets (i.e., interaction traces) to train a model and needs to combine the model with  
a set of pre-defined rules
to guide  testing. 

Reinforcement learning (RL) can teach machine
to decide which events to explore rather than relying on pre-defined
models or human-made strategies~\cite{hing2007reinforcement}. 
A Q-table is used to
record the reward of each event and the information
of previous testing knowledge. The reward function can be defined based
on the differences between pages~\cite{pan2020reinforcement} or unique activities~\cite{koroglu2018qbe}.
Reinforcement learners will learn to maximize cumulative reward with the goal of achieving higher code coverage or detecting more bugs.

While existing RL techniques have improved app testing, they 
focus on abstracting the information of app pages
and then use the abstracted features to train behavior models
for testing~\cite{borges2018guiding, vuong2019semantic}.  For example, QBE~\cite{borges2018guiding}, a Q-learning-based Android app testing tool, abstracts each app page into five categories
based on the number of widgets (e.g., too-few, few, moderate, many, too-many). The five categories are used to decide which events to explore. 
However, existing RL techniques do not understand the fine-grained
information of app pages like human testers normally do during testing,
such as the execution frequencies and the content of GUI widgets.
This is due to the limitation of the basic tabular setting of RL, which requires a finite number of states ~\cite{jin2018q}.
Therefore, the learned model may not capture the accurate behaviors of the app.
Also, many RL-based techniques focus on training each app independently~\cite{adamo2018reinforcement, vuong2019semantic, vuong2018reinforcement, mariani2012autoblacktest, pan2020reinforcement} 
and thus cannot transfer the model learned from one app to another.



To address the aforementioned challenges, we propose a novel approach,
\Name{}, based on deep Q-networks (DQN). 
\Name{}  learns a behavior model from a set of
existing apps and the learned model can be used to explore and 
generate tests for new apps. During the training process, \Name{}
is able to understand and learn the details of app events
by leveraging a deep neural network (DNN) model~\cite{larochelle2009exploring}. More precisely, 
we have developed a set of features taken as input by \Name{}. 
The insight of these features represents what a human tester
would do during the exploration. For example, a human tester may decide
which widget to execute based on its  content or how many times it was
executed in the past. \Name{} does not use any pre-defined rules or thresholds
to tune the parameters of these features
but let the DQN agent  learn a behavior model based
on the feature values (represented by vectors) automatically obtained during
training and testing phases. 

A key novel component of 
\Name{} is a deep neural network (DNN) model that can process multiple complex
features to predict Q value for each GUI event to guide Q-learning. 
With the DNN, \Name{} can be easily extended to 
handle other types of features.  Specifically, to test an app, \Name{} first trains
a set of existing apps to learn a behavior model. 
The DNN serves as an agent to compute  
Q values used to determine the action (i.e.,
which event to execute) 
at each iteration. 
In the meantime, \Name{} maintains a special event 
flow graph (EFG) to record
and update the feature vectors, which are used for DNN to compute 
Q values. 

Because the features are often shared 
among different apps, \Name{} is able to apply 
the model learned from existing apps to new AUTs. 
To do this, the agent continuously 
adapts the existing model to the new AUT 
by generating new actions to reach the
desired testing goal (e.g., code coverage). 

\ignore{ \yuc{Different Apps may share common features, for example, the visiting frequency of widgets, system event liking "back" and "restart", similar meaning texts on widgets, and so on. So \Name{} can transfer learning on new AUTs. If the Apps in the training set share more common features to the AUTs, the performance of transfer works better, for example, the Apps in the training set have buttons with text "OK"/"Film" and the AUT involves a widget with the same texts or synonyms "OK"/"Movie". We believe if there are sufficient Apps in the training set to provide enough samples, \Name{} can handle most possible cases for new AUTs. If different Apps do not share the common features, then there is no use to train a human tester about how to test an app.}
}


\ignore{
\Name{} differs from existing RL-based approach in several aspects.

The reinforcement learning reward is calculated

A key benefit of \Name{} is its extensibility. 

The key benefit of \Name{} is \commentty{mention extensibility.} \yuc{ \Name{} can be easily extended. It actually a structure which can handle multiple features with different types of format. Currently, we implement 3 features as content, visited times, and descendants on it. In the future, it can also handle feature as image. Simply saying, \Name{} uses deep neural network model. This model can handle multiple inputs with different types. If an UI information can be represented as a vector or a matrix, it can be inputted into \Name{} as a feature to guide the testing to achieve high code coverage. If the inputted feature can not bring any benefit for code coverage in the training, \Name{} will automatically discard it. The software interface of \Name{} is very simple, it can take multiple vectors and matrix as input. There are some default features pre-processing sub-model in \Name{} as matrix with 1D-CNN, and vector with full connected layer. User can also define their own pre-processing sub-model. For example, 2D-CNN for image and 1D-CNN for text. There is another extensibility for \Name{}. It likes all of the learning based method. The big training set \Name{} has visited and learned, the more powerful \Name{} is.} 

\Name{} framework is able to take into the all kinds of 
features (e.g., GUI content, execution frequency,
an event flow graph) of an app page like a human tester.

\Name{} differs from existing RL-based approach in two main aspects. 
First, 
\Name{} also utilizes reinforcement learning to provide testing strategy about which event to trigger. 
Comparing to existing methods, \Name{}  can input text, image, execution frequency, event flow tree as the feature 
and let reinforcement learning to understand the meaning of them and 
make an event decision based on them. 

Unlike existing Q-learning, where the model
is learned based on the differences between pages or 

}

In summary, our paper makes the following contributions:

 \begin{itemize}[noitemsep,nolistsep]

 \item 
 An approach to testing Android apps based on deep
 Q-learning.  
 
 \item
 A novel and the first deep Q-learning model that can process
 complex features at a fine-grained level.

\item 
 An empirical study showing that the approach is effective at achieving higher
 code coverage and better bug detection than the state-of-the-art tools. 
 
 \item
The implementation of the approach as a publicly available tool,
\Name{}, along with all experiment data~\cite{DinoDroid}.

\end{itemize}

\ignore{The reminder of this paper is organized as follows. 
We present motivation and background  in Section~\ref{sec:motivation}. 
Section~\ref{sec:approach} describes the approach of \Name{}, followed
by the evaluation~\ref{sec:results}. 
We then discuss the limitations and related work in Section~\ref{sec:discussion} and \ref{sec:related},
and finally conclude our work in Section~\ref{sec:conclusion}.
}

\ignore{
Reinforcement learning technologies are used to test android application too. 
Comparing to heuristic method, reinforcement learning based testing tool can 
make a testing strategy by a machine rather than a directly from human experience. 
In the reinforcement learning tools, page information and widget information 
on app page are represented as features. 
Reward are designed for target. For example, reward can be visiting more unsimilar pages[][] or 
getting highest code coverage(this paper). Reinforcement learning learn a strategy how to 
decide which event to trigger based on the event feature and exploration reward. 
Then it can achieve the highest value of target in the testing. 
The advantage of reinforcement learning is machine can summarize testing 
rules in training set. If an app needs testing strategy to handle with, 
reinforcement learning can provide a new strategy based on it with training 
rather than human modify a existing strategy to match with it. 
By the way, reinforcement learning beats heuristic method and even 
for human in some game competitions like Alphgo and Alphastar 
which means reinforcement learning can also provide competitive strategy with heuristic method.

There are two types Reinforcement learning android testing tools(From ISSTA paper's related works). The first type of reinforcement learning tools model app independently. Most of these tools extend  AutoBlackTest [35] (ISSTA says they do not have training process, we can discuss how to organize it ). In every app, it trains and tests the app at same time with a Q-table. Every unique event is a state in Q-table. If an app has tons of events, the Q-table can not provide so many states. Because of in the Q-table every event has its specific identifier, this kind of testing can not do transferring between app with training and testing. Another type of reinforcement learning has an
explicit training process to learn from the previous testing process and the learned experience will later be leveraged on new apps. Because of the needed of transfer, they abstract the feature of events to simple representations as event number with S = \{too-few; few; moderate; many; too-many\} to fill a Q table. Lots of necessary information are loss in this abstract process and the tool can not get a good result.

The ideal reinforcement learning android testing can achieve the advantages of both of methods. It should like a human to test app. A human may learn to summarize knowledge about how to test an app based on testing for tons of apps. Then he can be an expert to test a new app. In the testing process that human can still get knowledge from new app and adapt his experience on it. The most important thing is, human can see and get understand the detailed information as text of the app in this process, not only the abstract representation. However, it is not easy. As said in (deep learning paper ase), "Because it is hard for a machine to understand the content within a GUI element, it is also difficult to determine which button to click". In this paper, we approach a AAA which can like a real human, get knowledge on some apps and adapt it to test on a new one. In this process, AAA can also see clear the information on every event and understand it. For example, it can understand the image meaning, text meaning, and even for the model on every event.

}



\section{Motivation and Background}
\label{sec:motivation}

In this section, we first describe a motivating example of \Name{},  followed
by the background of deep Q-networks (DQN), the problem formulation,  and the discussion of
existing work. 

\subsection{A Motivating Example.}

\begin{figure}[t]
\vspace*{-5pt}
\centering
\scalebox{0.77}{
\includegraphics[scale=0.5]{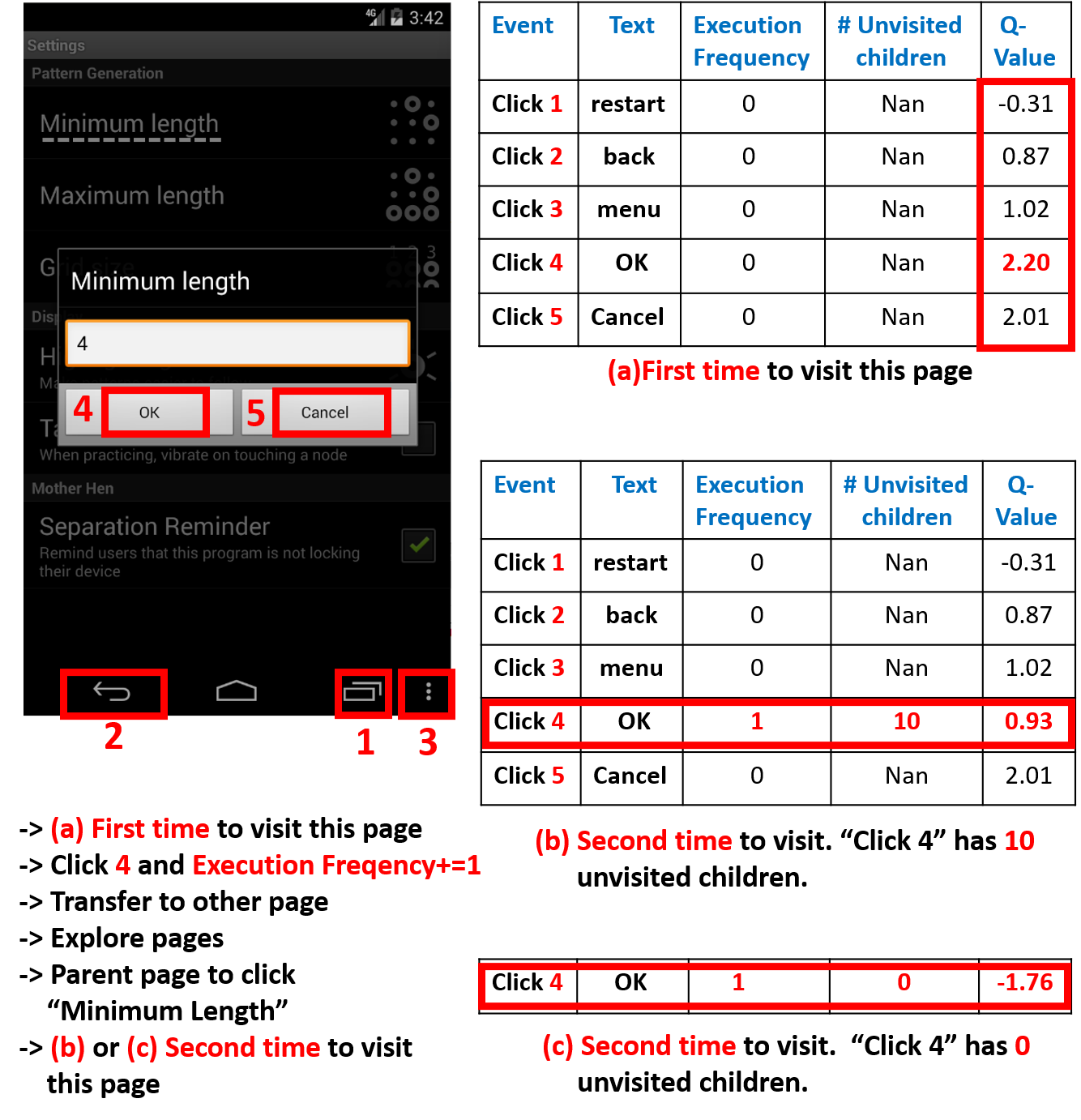}
}
\vspace*{-15pt}
\caption{A Motivating Example }
\label{fig:tree}
\vspace*{-15pt}
\end{figure}

Fig.~\ref{fig:tree} shows an example of the app  \textit{lockpatterngenerator}~\cite{lockpatterngenerator}. 
\yuc{This simple example demonstrates the ideas of DinoDroid,
but the real testing process is much more complex.
}
After clicking ``Minimum length", a message box pops up with a textfield and 
two clickable buttons. Therefore, the current page of the app has a total of five events (i.e., ``restart", ``back", ``menu", ``OK", and ``Cancel"). 
The home button is not considered because it 
is not specific to the app. 
When a human tester encounters this page,  he/she needs to decide 
which event to execute based on his/her
prior experience. 
For example, a tester is likely
to execute the events that have never been executed before. 
\yuc{The tester may also need to know the execution context of
the current page (e.g., the layout of next page) 
to decide which widget to exercise. 
}


In this example, suppose none of the five events on the current page have been executed before. Intuitively, the tester tends to select the ``OK" event to execute because it is more likely to bring the app to a new page. ``Cancel" is very possible to be the next event to consider because ``restart", ``back", and ``menu" are general events, so the tester may have already had experience in executing them when testing other apps. 
In summary, to decide whether an event has a higher priority
to be executed, the tester may need to consider its ``features", such 
as how many times it was executed (i.e., execution frequency)
and the content of the widget. 
\Name{} is able to automatically learn a behavior model from a set of
existing apps based on these features and the learned model can be used 
to test new apps.

Tab.(a)-Tab.(c) in Fig.~\ref{fig:tree} are used to illustrate
\Name{}.  In this example, 
\Name{} dynamically records the \emph{feature values}
for each event, including the 
execution frequency, the number of 
events not executed 
on the next page (i.e., child page), and the text
on the event associated with the event. 
%
Next, \Name{} employs a deep neural network to predict the accumulative reward 
(i.e., Q value)
of each event on the current page based on the aforementioned features and
selects the event that has the largest Q value to execute. 

Tab.(a) shows the feature values and  Q values when  
the page appears the first time. Since ``OK" has the largest Q value, it is 
selected for execution. 
 \Name{}  continues exploring the events on the new page
and updating the Q value.  When the second time this page appears, 
the Q value of the event on executing ``OK" button decreases because it is already 
executed.  
As a result, ``Cancel" has the largest Q value and is selected for execution. 
In this case (Tab.(b)),
the child page of ``OK" contains 10 unexecuted events. However, suppose
the child page contains zero unexecuted events (Tab.(c)),  the Q value 
becomes much smaller. This is because \Name{} tends to select the event
whose child page contains more unexecuted events. 


\yuc{The underlying assumption of our approach is that 
the application under test should follow the principle of least surprise (PLS). If an app does not meet the PLS, e.g., an ``OK" textual widget is incorrectly associated
with the functionally of ``Cancel", it would mislead
\Name{} when finding the right events to execute. Specifically, \Name{} exploits the learned knowledge to execute correct events that result in higher code coverage or triggering bugs.}



%


\subsection{Background}
\label{sec:introduction}

\subsubsection{Q-Learning}

Q-learning~\cite{watkins1992q} is a model-free reinforcement learning method which seeks to learn a behavior policy for any finite Markov decision process (FMDP), Q-learning finds an optimal policy, $\pi$, that maximizes expected cumulative reward over a sequence of actions taken. Q-learning is based on trial-and-error learning in which an agent interacts with its environment and assigns utility estimates known as Q values to each state. 

\begin{figure}[t!]
\centering
\scalebox{0.8}{
\includegraphics[scale=0.55]{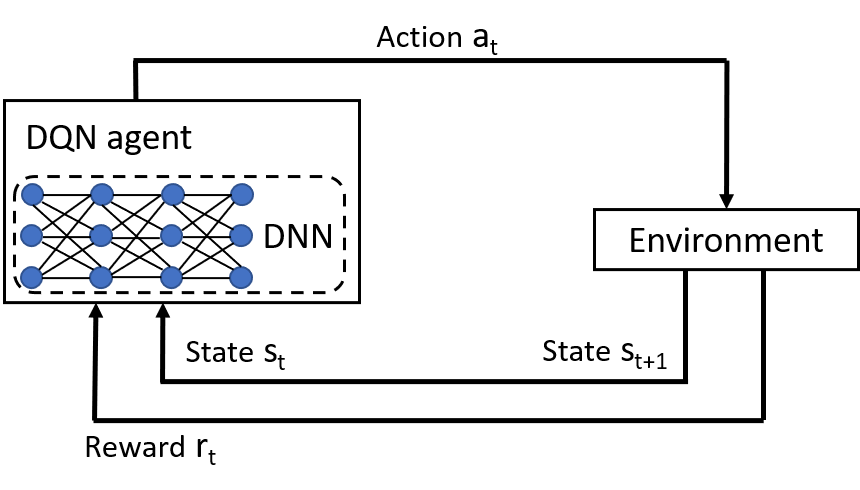}
}
\vspace*{-10pt}
\caption{Deep Q-Networks}
\label{fig:dqn}
\vspace*{-10pt}
\end{figure}

As shown in Fig.~\ref{fig:dqn} the agent iteratively interacts with the outside environment. 
At each  iteration $t$, the agent selects an action $a_t \in A$ based on the current state $s_t \in S$ and executes it 
on the outside environment. After exercising the action, there is a new 
state $s_{t+1} \in S$, which can be observed by the agent.
In the meantime, an immediate reward $r_t \in R$ is received. 
Then the agent will update the Q values using the Bellman equation~\cite{bellman1952theory} as follows:

\begin{center}
$Q(s_t,a_t) \leftarrow  Q(s_t,a_t)+\alpha*(r_t+\gamma*\underset {a} \max Q(s_{t+1},a)-Q(s_t,a_t))$
\end{center}

In this equation, $\alpha$ is a learning rate between 0 and 1, $\gamma$ is a discount factor between 0 and 1, $s_t$ is the state at time $t$, and $a_t$ is the action taken at time $t$. Once learned, these Q values can be used to determine optimal behavior in each state by selecting action $a_t= \mathop{\arg\max}_{a} Q(s_t,a)$.

\subsubsection{Deep Q-Networks}
\label{subsubsec:dql}

Deep Q-networks (DQN) are used to scale the classic Q-learning to more complex state and action spaces ~\cite{mnih2015human, arulkumaran2017deep}. For the classical Q-learning, $Q(s_t,a_t)$ are stored and visited in a Q table. It can only handle the fully observed, low-dimensional state and action space. As shown in Fig.~\ref{fig:dqn}, in 
DQN,  a deep neural network (DNN), specifically involving convolutional neural networks (CNN) ~\cite{Conv1D}, is a multi-layered neural network that for a given state $s_t$ outputs Q values for each action $Q(s_t, a)$. Because a neural network can input and output high-dimensional state and action space, DQN has an ability to scale more complex state and action spaces. A neural network can also generalize Q values to unseen states, which is not possible when using a Q-table. It utilizes the follow loss function~\cite{arulkumaran2017deep} to alter the network to minimize the temporal difference (TD)~\cite{dayan1993improving} error as a loss function $loss=r_t+\gamma*\underset {a} \max Q(s_{t+1},a)-Q(s_t,a_t)$. The $\gamma$ is the discount factor which is between 0 and 1. In other word, with the input of $(s_t,a_t)$, the neural network is trained to predict the Q value as:

\vspace*{-10pt} 
\begin{equation}
Q(s_t,a_t)= r_t+\gamma*\underset {a} \max Q(s_{t+1},a)
\label{equ:1}
\end{equation} 
\vspace*{-15pt} 
 
\ignore{
\begin{center}
$Q(s_t,a_t)= r_t+\gamma*\underset {a} \max Q(s_{t+1},a)$					(1)
\end{center}
}

So in a training sample, the input is $(s_t,a_t)$ and output is the corresponding Q value which can be computed by $r_t+\gamma*\underset {a} \max Q(s_{t+1},a)$.

\ignore{
 We can compute the Q value $Q(s_t,a_t)= r_t+\gamma*\underset {a} \max Q(s_{t+1},a)$ (1). Then providing $r_t+max_a(Q(s_t+1,a))$ and $Q(s_t,a_t)$.
}

\subsection{Terminologies}


A GUI  \textit{widget} is a graphical element of an app, such 
as a button, a  text field, and a check box. 
An \textit{event} is an executable GUI widget
with a particular event type (e.g. click, long-click, swipe, 
edit), so a widget can be associated with one or more events.  
In our setting, 
a \textit{state} $s$ represents an app page (i.e., a set of widgets shown on the current screen. If the set of widgets is different, we have another page). 
We use $s_t$ to represent the current state and $s_{t+1}$ to represent the next state. 
A \textit{reward} $r$ is calculated based on the improvement of coverage. 
If  code coverage increases, $r$ is assigned a positive number 
($r$=5 by default); otherwise,  $r$ is assigned a negative number ($r$=-2 by default). 
An \textit{Agent}  chooses which event to execute 
based on the accumulative rewards (i.e., Q values)
on the current observed state. 
The chosen event is an \textit{action}. 
For example, the agent performs an action by 
choosing to ``click a button A" on the current page. 
A \textit{Policy} is $\pi(a,s)=Pr(a_t=a|s_t=s)$, which maximizes the expected cumulative reward.
Q-learning learns a policy to tell the agent what action to take.

\ignore{\yuc{Users can access widgets to interact with android apps. A widget can be a button, clickable text, clickable image, checkbox, and so on. An \textit{event} $e$ is the exercise of a GUI widget with a particular action type
e.g., click/long click "search" button. So one "search" button may have two events: 1. click "search" button. 2. long click "search" button. An \textit{action} $a$ is an event issued by \Name{} to exercise a GUI widget. It is an Q learning term which is often related with Q learning discussion in the reminder of this paper. After agent selects an event $e$, it performs action $a$ to exercise the event $e$ on the page.} 
}

\ignore{

\subsection{Human Model}

The key idea of \Name{} is generating event sequences guided
by the behaviors of human testers.  We have defined 
three types of human behaviors that can be input as features. 

\begin{itemize}
\item \textbf{Frequency of visited events:} 
Human testers want to make sure that every GUI component gets a chance to be exercised
as its as visiting frequencies. \commentty{some events are naturally visited more often
than others...}

\item \textbf{Types of events:} 
Different types of events are not equally selected. For example,
a study~\cite{su2017guided} has found that \commentty{which type
of events could be more important?}

\item \textbf{Number of unvisited children events:} 
Human testers tend to select events that are likely 
to bring the app to a new page. 

\end{itemize}

}


\ignore{
When a human tester are testing android UI, he may use his before knowledge to guide his behavior to test it. The authors in Stoat ~\cite{su2017guided} summarize these human behaviors. They investigated 50 most popular Google Play apps from 10 categories(e.f., Education, Business, Tools). On each app, they manually explored as many functionalities as possible. Three observations are summarized based on their experience based on these 50 apps.

1. Visited Frequency. They find all events should be given chance to be executed. Every time, the event  with lowest Visited Frequency in the current page should be provided as higher priorities to execute.

2. Type of Events. In the 50 apps, they find different types of events are not equally selected. It should gives different priorities to execute different types of events such as normal UI events(e.g. click), navigation events(e.g. back, scroll, and menu).

3. Number of Unvisited Children Events. The events solicits more new UI widgets on the next page should be given higher priorities to execute. Because they may reach new functionalities in behind pages.
}

\subsection{Limitations of  Existing Q-Learning Techniques }
\label{subsec:limitations}

\ignore{
While there have been techniques on modeling human-like
features to improve testing~\cite{?}, they rely on manually
defined heuristics.  For example, Stoat~\cite{su2017guided}, a model-based testing 
tool,  encodes the features (parameters) in the model and uses
them to guide the exploration. However,  values of
the parameters are tuned by a list of pre-defined
heuristics.  \commentty{Need an example.}. The limitations 
are \commentty{what would be the problem.}

\yuc{For example, there is a execution weight function in stoat. The event with higher execution weight will be assign higher priority to execute. In that function, there are three parameters $\alpha$, $\beta$,$\gamma$ to multiply with features. These parameters are tuned during human investigation on the 50 Google Play apps. There are two limitations for their approach. 1. They can not design complex features. All of features in stoat are single values, such as visited times. They can not handle complex features as text which has more dimensions. 2. The number of features are limited. Because each feature needs a parameter to tune. Parameters on different features need to work cooperatively. Every parameter will add a new degree of freedom of configuration. To make a good configuration to achieve a good performance is not easy. 3. The man heuristic method with parameters need professional people to tune. The performance of the system is due to the ability level of the tuner. The tuner's ability may have a up limitation. Recently, AI wins human in plenty of areas. 4. The tuned parameters are not extendable. For example, after investigating 50 more new Google Play apps, the tuner may change his mind about the heuristic design or parameters configuration. }

Existing RL-based Android testing techniques 
focus on Q-table. \commentty{Elaborate}
}

The techniques that are mostly related to \Name{} are 
Q-learning-based Android app testing~\cite{adamo2018reinforcement, vuong2018reinforcement, mariani2012autoblacktest, pan2020reinforcement}.  
These techniques are all based 
on a Q-table, which have several limitations. 
First, Q-table is not capable of handling high-dimensional features, such as text and image --- adding a new feature space for a state to the Q-table will lead to
its exponential growth.  For example, a word embedding feature,
represented as an M-dimension vector with $N$ possible values, 
would need $O(m^n)$ columns in a Q-table. Second, 
most existing techniques use the resource ID of an event
to represent the state in a Q-table. However, different apps can have 
different ID assignments. Therefore, these techniques
often focus 
on training and testing individual apps and cannot train a model
from multiple apps to test new apps.
The only work that uses Q-table  to transfer 
knowledge among apps is QBE~\cite{koroglu2018qbe}. 
In order to limit the size of Q-table, QBE 
abstracts the state (i.e., an app page) into 
five categories in terms of the number of 
widgets on the current page and
actions into seven general categories
(i.e., menu, back, click, long-click, text, swipe, contextual).
However,  such abstractions may cause
the learning to lose a lot of important information when
making decisions on which events to execute.


\ignore{instead of using resource id to represent
each state, QBE abstracts the page information into five states:
too-few, few, moderate, many, and too-many, based on the 
number of widgets.}

Unlike traditional Q-learning, \Name{} designs
a novel deep neural network (DNN) model 
that can handle complex features with infinite feature space,
such as word embedding with n-length vector and 
high-dimensional image matrix. 
\Name{} is able to process these features at a
fine-grained (ie., widget) level as opposed to 
abstraction~\cite{koroglu2018qbe}.  \Name{}'s features 
are specifically designed to be general across apps,
so the knowledge learned from existing apps can be transferred 
into new apps. For example,  the feature regarding ``execution frequency"
applies to almost every app because exploring new events
are always desirable.  Also, when considering the ``textual content"
of events,  an ``OK" button in every app may have similar meaning
across different apps.


\ignore{\commentty{Need to discuss ``a simple DQN with limited functions ~\cite{vuong2019semantic}."}
\yuc{QDROID ~\cite{vuong2019semantic} uses Deep Q learning to learn the behavior of android exploration. However, its neural network in the agent is too simple to handle complex features. Actually, it can only input a vector of length 4. This vector abstracts the page information into 4 groups as as "input", "Navigation", "List", and "Button". Every item in the vector is number of widgets belong to each group. The output of the neural network is also a vector of length 4 which indicates the probability distribution that for item in the group as "input", "Navigation", "List", and "Button". For example, if the DQN
outputs the probability distribution for (Input, Navigation,
List, Button) respectively as (0.6, 0.2, 0.1, 0.1), the Planner
will look for a component of group Input first It only provides abstract action the rather than which event to execute. The missing information in the observation and action can not let a agent learn a good behavior to test android app.}
}

\ignore{
\yuc{In existing works, they all use classic Q-learning with Q table ~~\cite{adamo2018reinforcement, vuong2018reinforcement, mariani2012autoblacktest, pan2020reinforcement} or combining Q-table with a simple DQN with limited functions ~\cite{vuong2019semantic}. Every new feature space for a state will bring  Exponential Growth space requirement for a Q-table. A feature as word embedding which can be represent to n dimensions vector with m possible values needs $O(m^n)$ rows of Q table. It is impossible stored and process in a table. So they can not handle high dimensional states which does an android app GUI have as image and text. Actually, most of the existing works utilize the unique system id of event as the state of Q table, because the id space is 1 which is easy to handle by Q-table. There are two additional limitations besides no complex feature. 1. Sometimes, an complex android App may have a big number of different events as state. In the Q-learning, the Q value from the subsequent state-action pair is propagated to affect to preceding pairs. So it needs to visit every event many times to update its Q value which can be influenced by its children event Q value. So it needs big amount of time to converge Q table to be precise in the Android testing. 

2. Different app have different system ids assignment. Then the Q-table trained on one App can not be used on any other app, even for the different release version of the same App. There is only one existing work QBE~\cite{koroglu2018qbe} uses Q-table and transfer knowledge among different apps. However, it pay a huge cost to achieve that because of using of Q-table. Does not like other works, QBE does not use resource id to represent every state. It abstracts the page information in deep compression state. The abstract states are S = too-few; few; moderate;many; too-many. The abstract actions are A = menu; back; click; longclick; text; swipe; contextualg. After that, they can use a limited space’s table to store value crossing between a state and an action. The machine is designed to choose an abstract action based on an abstract features. Because QBE abstracts all page's information into a number with 7 possible value. Q-table can use 7 rows to represent it. However, most of page information are lost in this abstract process. And the action is also an abstract. It can only claim the action type but no information about which event to click.}

Unlike model-based testing, \Name{} can automatically learn the
the information of the features as previous experience. 
The increase of code coverage 
Unlike other RL-based Android testing, \Name{} designs
a novel  neural network model to save the feature information
in the previous explorations. 
This model can handle complex features with infinite feature space. 
\yuc{It utilizes deep neural network to store and read the Q value with infinite feature space. Multiple complex features with different types can be input into the deep neural network to obtain the Q value. For example, this deep neural network can input the text embedding with n length vector and image matrix with m*m are the same time. The feature's space and number are not restricted in \Name{}. By doing this, \Name{} can see what a human can see and learn what a human can learn. Because different apps also share some common knowledge as influence of visited time of event or meaning of text as "OK". Because of this, \Name{} is able to apply the learned model to a new app and adapt the model on it. }

\commentty{May need to say more.}
Because of this, \Name{} is able to apply the learned model
to a new app. 
}

\ignore{
\subsection{Background}

All above 3 features are never used to train behavior in the learning based model. Text meaning feature are never used by all of existing android testing works. Why AAA can like a human to understand and learn behavior from the text and other complex features and all existing works can not do it. It is because AAA utilizes a novel model structure the existing works do not utilize them. 

The existing works and AAA both utilizes Q-learning Bellman equation to learn the strategy about how to test android app to increase the code coverage. $Q^{new}(s_t,a_t) \leftarrow  Q(s_t,a_t)+\alpha*(r_t+\gamma*\underset {a} \max Q(s_{t+1},a)-Q(s_t,a_t))$ (1). In the equation, $s_t$ is the state and $a_t$ is the action at the time step t(ISSTA 2020 paper has a detailed introduction). $\alpha$ is a learning rate and $\gamma$ is a discount factor. The idea of Q-learning is using time t's observed reward $r_t$ and next time step's t+1's maximum Q value to update current Q value. Based on this equation, how to make a decision to choose an action in step t is not only related time step t but also related with time step t+1. The decision equation is $a_t= \mathop{\arg\max}_{a} Q(s_t,a)$. Because the current $Q(s_t,a)$ has stored the $s_t+1$ information in equation(1), Q-learning's decision in $s_t$ also put possible next state's information into consideration.

Now the problem is how to store the $Q(s_t,a_t)$. In the existing method likes QBE ~\cite{koroglu2018qbe}, they all utilize Q-table to store the state as in Fig 2. Because of Q-table's limitation, they can not use it to store infinite states. There are two solutions for them to deal with that. In the transformable tools like QBE ~\cite{koroglu2018qbe}, they compress the features be abstract features. For example, they compress a page feature to the number of enable actions as S = \{too-few; few; moderate; many; too-many\}. They also compress the action's field to A =\{ menu; back; click; longclick; text; swipe; contextual\}. After that, they can use a limited space's table to store value crossing between a state and an action. The machine is designed to choose an abstract action based on an abstract features. It can not see the detailed information on screen like a human tester. So it will not has a high performance. Another solutions store every state as event identifiers which are unique ids in one app. This kind of solutions can not store many features other than 
the identifiers because the table space is limited. The large table space will bring problems at the training. The most important thing is it satisfy the model transform among apps, because the unique ids only be available in one app. They can not transform to others.

AAA's technology are different with their two. AAA utilizes a structure as in Fig. 2 to store the state, action and Q value. The states in AAA are event's features which can mach actions. Every states in AAA can only match to one action with a Q value. The state's Q value is stored in a neural network as a Deep Q learning idea. Because a neural network can handle infinite feature space as input, so AAA has an ability to understand any kind of features even for the natural language text and do an action based on them. In every page, AAA will utilize this structure to process features of each events to compute a Q value. Then it selects the event with biggest Q value to do action. In the update of the model, AAA computes all actions' Q value in $s_{t+1}$ based on the neural network structure to get the maximum the Q value combined with the $r_t$ to update the $Q(s_t,a_t)$. As in Fig. 2, AAA's neural network utilizes a unique sub-neural network to pre-process each inputs. The nlp text is handled by a CNN network which is a common solution to deal with nlp text. The visited time and event flow graph are handled with dense networks. In the feature, other new features will be suggested to use specific feature to handle. As we known, CNN is also good to pre-process image of screen shot of events. Then it combine the pre-process results of each input as one vector. Then a new neural network is designed to process the vector and gives a result as the Q value.
}

\section{\Name{} Approach}
\label{sec:approach}

\begin{figure}[t]
\vspace*{-5pt}
\centering
\scalebox{0.8}{
\includegraphics[scale=0.55]{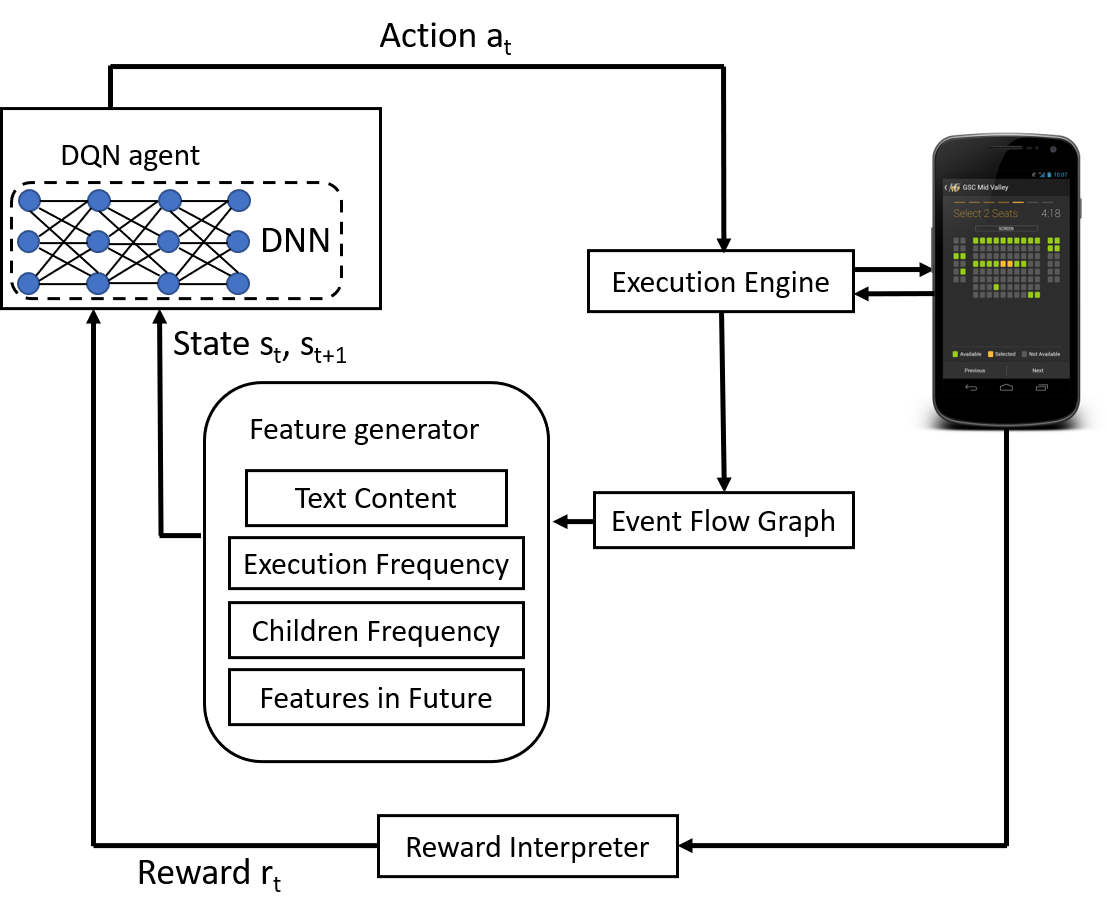}
}
\vspace*{-10pt}
\caption{Approach Overview }
\label{fig:overview}
\end{figure}

Fig.~\ref{fig:overview} shows the overview of \Name{}. 
 %
An iteration $t$ begins with an app page.  
In the current state $s_t$, \Name{} selects the event $e_t$ with the highest
accumulative reward (i.e., Q value), performs the corresponding action $a_t$, 
and brings the app to a new state $s_{t+1}$. 
After performing $a_t$,  \Name{} employs a special
event flow graph to generate feature values for each
event in $s_{t+1}$. 
A reward $r_t$ is generated based on the observed code coverage and app crashes.
A positive number is assigned to $r_t$ if code coverage increases
or a unique bug is detected and a negative number is assigned
if the coverage does not change. 
The  deep Q-network
agent uses a neural network model to compute and update the Q value for the event $e_t$ responding with $a_t$. The learning process continues iteratively from the apps provided as the training set.  When testing a new app, \Name{} uses the learned model as the initial model to guide the testing. The model is updated following the same process until  a time limit is reached or the coverage reaches a plateau.
 
\ignore{
 Fig.~\ref{fig:overview} shows the overview of \Name{}. 
 %
An iteration $t$ begins with an app page.  
\Name{} selects the event $e_t$ with the highest
accumulative reward as Q value, performs the corresponding action ($a_t$), 
and brings the app to a new state ($s_{t+1}$). 
After exercising $a_t$,  \Name{} employs a special
event flow graph (EFG) to generate feature vectors for each
event in $s_{t+1}$. The DQN agent uses a equation(1) to compute the Q value for the $e_t$ responding with $a_t$ . A neural network is used to learn the policy as $e_t \in s_t$ owns the accumulative reward as Q.
The learning process continues iteratively from the apps provided
as the training set.  When testing a new app, \Name{} uses
the learned model as the initial model to guide
the testing of the new app. The model is updated following
the same process until it is adapted to the new app (e.g.,
a time limit is reached or the coverage reaches a plateau).
} 
 
 \ignore{
 A positive reward is provided if the coverage increases, otherwise,
\Name{}  issues a negative reward. The execution engine records
the information of the new page and sends it to
the agent for generating an event action for the next iteration. 
\Name{} employs an event flow graph (EFG) to record the 
current state $s_{t}$.  The EFG is updated for each iteration when a new
event is exercised.  Within the learning agent, a deep neural network is used
to predict Q values (equation (1)) of state  $s_{t}$ given a set of features
and the previous state $s_{t}$. By default, \Name{} considers  three features:
content of GUI events, visiting frequency, and page context. Other features
can be easily integrated.  
The event with the largest Q value will be chosen to send to the execution engine 
as the next action and perform the next iteration. 
}

\setlength{\textfloatsep}{0pt}
\begin{algorithm}[tb]
\scriptsize
\caption{\textbf{\small{\Name{}'s testing}}}
\label{alg:agent}
\begin{algorithmic}[1]
\raggedright
\Require App under test $AUT$, DQN's Model $M$ with before knowledge, execution time LIMIT
\Ensure  updated new $M$
\If{$M$ not exist}
\State $M$ $\gets$ \texttt{buildNewModel}() /*First time to run*/
\EndIf
\State Memory $\gets$ \o{} /*Memory stores previous samples */
\State G $\gets$ \o{} /*Initialize event flow graph */

\State $p_0$ $\gets$ \texttt{execute}(AUT) /*First lanuch to get the first page $p_0$*/

\State G $\gets$ \texttt{updateGraph}($p_0$, G) /*update G with new page*/

\State $s_0$ $\gets$ \texttt{featureGenerator}($p_0$,G) /*Every event in $s_0$ contains 3 features*/

\While{t $<$ LIMIT}

\If{$t$ mod 10 equals 0}
\State sendSystemEvent() /*send random system event*/
\EndIf
\State $a_t$ $\gets$ \texttt{getActionEvent}($s_t$, M) /*Event selection */
\State $p_{t+1}$, codeCoverage, crash $\gets$ \texttt{execute}(AUT, $a_t$)
\State $r_t$ $\gets$ \texttt{rewardInterpreter}(codeCoverage, crash)
\State G $\gets$ \texttt{updateGraph}($p_{t+1}$, G) /*update G with new page*/
\State $s_{t+1}$ $\gets$ \texttt{featureGenerator}($p_{t+1}$,G) 
\State $Q(s_t, a_t)=r_t+\gamma*\underset {a} \max Q(s_{t+1},a)$   where $\gamma=0.6$
\State $Q$ $\gets$ $Q(s_t, a_t)$
\State batch $\gets$ \texttt{extractTrainBatch}(Memory) $\mathop{\cup}$ ($a_t$,$Q$)
\State M $\gets$ \texttt{updateModel}(batch, M) /*Learning for DQN */
\State Memory $\gets$ Memory $\mathop{\cup}$ ($a_t$,$Q$)
\State $s_{t+1}$ $\gets$ $s_t$
\EndWhile
\State return M
\end{algorithmic}
\label{alg:overview}
\end{algorithm}

\subsection{\Name{}'s Algorithm}

Alg.~\ref{alg:overview} shows the details of \Name{}. 
\yuc{When testing or training an app, \Name{} first checks
whether there exists a learned model,
which saves  
all learned knowledge 
(weights on neural networks) of \Name{}.}
If so, the model will be updated
to adapt this app; otherwise, \Name{}
starts building a new model (Line 2).
A memory is used to record the samples from earlier iterations (Line 3),
where each sample contains the feature values of the executed
event and the associated Q value.
An event flow graph (EFG) is initialized for each app at the beginning
of training or testing (Line 4).  
\yuc{Note that the memory and the EFG will
be abandoned after completing an 
iteration of the current app because the needed information has already been learned and saved in the model.
}
The details of the EFG
will be described in Section~\ref{subsubsec:efg}. 
Next,
\Name{} launches the app and reaches the first
page. The initial state of DQN is obtained 
and the EFG is updated accordingly (Lines 5--6).
The algorithm then takes the current page and the EFG to generate 
features for each GUI event on the current page (Line 7). 
\Name{} considers three types of features as described in Section~\ref{subsec:feature}.

Now the iteration begins until a time limit is reached (Lines 8--22).
To amplify the chance of bug detection, \Name{} issues 
a random system event at every 10 iterations (Line 9--10) by using Stoat \cite{su2017guided}'s system event generation library. 
At each iteration, \Name{} uses a deep neural network 
model to select the event with the highest
accumulative reward (i.e., Q value)  and perform the corresponding
action (Line 11).
The details of the getActionEvent will be described in Section~\ref{subsubsec:event}.   
After the execution, the algorithm obtains three kinds of information: the new page, the
current code coverage, and the crash log (may be empty) (Line 12). 
The information is used to compute the reward $r_t$ (Line 13). 
Then, the event flow graph G is updated based on the new page $p_{t+1}$ (Line 14). 
With the updated G and the new page, the algorithm can generate
features for each event in the state $s_{t+1}$ (Line 15).  
Based on $s_t$, $s_{t+1}$, $a_t$, and $r_t$, \Name{} uses the equation (1) 
in Section~\ref{subsubsec:dql} to compute the Q value of each event
in $s_t$ (Line 16).
To train the neural network, it uses a set of training samples, including both
the current sample and history samples.  
Each sample uses $a_t$ as input and Q value as output (i.e., label). 
The history samples (obtained from earlier iterations)
are recorded in a memory (Lines 17--20). \yuc{In each iteration, the history sample size is set to be 4 due to cost-effectiveness. A larger history sample size can be set by users to retain more
historical information for training, but it may also 
lead to a much higher cost.}

\ignore{
Then feature generator will generate features as $s_{t+1}$ which includes text embedding, visited frequency, and neighbor frequency, and other features which can be designed in the future version. At the same time, the reward interpreter checks whether this new event action result in a code coverage increase. The reward interpreter provides a positive reward $r_t$ as 5 to represent the code coverage increase. It uses a negative reward $r_t$ as -2 to represent the code coverage no increase. Deep Q Learning Agent gets $s_{t+1}$ from  Event graph flow generator and the $r_t$ from the reward interpreter. With the stored $s_{t}$ and $s_a$ from the last circle, Deep Q Learning can utilize the equation (1) to update the Deep neural networks. After that, it utilizes that network to predict the Q value of each event in $s_{t+1}$. The event with biggest Q value will be chosen to send to Uiautomator as the next action and do circle again. All of these operations are same at the training and testing. In the training, Deep Q Learning can continue to learn knowledge from apps in the training set based on these circles. In the testing, Deep Q learning can use the neural network's model from learned training as the before experience. Maybe the new app in the testing set can not perfect match the before trained knowledge. Then it can adapt the before model to the apps by updating the before model in the exploration.
}

\ignore{
\subsection{The \Name{}'s Overall Algorithm}

Alg.~\ref{alg:agent} describes the overall algorithm of  \Name{}. 
The algorithm takes as input the model $M$ learned from previous iterations,
\commentty{what does an iteration refer to?}
the app under training/testing, and time limit. 
\Name{} first checks if the model exists. \commentty{what does ``exists" mean?}
If not, \Name{} will use DQN to learn current app and generate a model (Section~\ref{??}). 
Otherwise, \Name{} will update the model with the new app (Line 3--Line 23). 

Specifically, \Name{} builds an event flow graph to save the
context the app exploration. 
\commentty{not sure if the following text is needed.}
\Name{} would build a reply set of every app. In the training process, 
\Name{} would select some samples to train the neural network based the sample balance policy. 
The memory on the line 4 is used to store the replay set. At the beginning, at the first start page of App, 
\Name{} utilizes Uiautomator to capture the page information as line 5. One the line 6, the graph generator 
will analysis and transfer the current page $P_t$ to a table $tab_t$ with method discussed in 
before section.  \commentty{Which section?}
Then \Name{} also update the Graph G by adding the $tab_t$. 
After \Name{} starts the circle till the timeout.

Under each iteration, the algorithm extracts features from the current table $tab_t$ and G (Line 8, Section~\ref{?}).
The features are encoded into state $s_t$, where the sub-states represent the events
in the current page ($e_1$,$e_2$,...$e_n$). 
The algorithm then uses getActionEvent() to decide which event should  be executed (Line 9, Section~\ref{?}).
\commentty{Is it from Q table?}.
The execution of the new event will bring the app into a new page 
and the code coverage is detected. 
Next, the algorithm assigns a reward $r_t$ to DQN depending on whether
the code coverage is increased or not. 
Based on the new page, \Name{} will map it to the table $Tab_{t+1}$ and update the Graph and the next state
$s_{t+1}$ is constructed.

\commentty{what about testing a new app? Can we simply the algorithm?}

}

\subsection{Feature Generation}
\label{subsec:feature}

\Name{} generates features for each event $e$.
These features are fed into a neural network as shown in Fig. \ref{fig:model}
so that the agent can choose an action 
(i.e., which event to exercise). \Name{} currently supports three features,
namely, \textit{execution frequency of current events}, 
 \textit{execution frequency of children events},  and \textit{textual content of events}. 
\Name{} employs an event flow graph (EFG) at runtime 
to obtain the features and transforms them into numerical data format vectors
provided as inputs to the neural network. 



\subsubsection{Types of Features}
\label{subsubsec:featureType}

\subB{Execution frequency of current events (FCR).}
The insight behind this feature is that 
a human often avoids repeatedly executing an event that brings
the app to the same page.   
Therefore, all GUI events should be given chances to execute. 
Instead of using  pre-defined rules or thresholds to weigh different events to decide which
ones should be assigned higher priorities to execute~\cite{su2017guided}, 
\Name{} automatically guides the selection of events
based on the recorded FCR feature value. 
FCR encodes the number of times that each event was executed 
during the exploration. 
The feature value is updated by adding one when
the event is executed.

\ignore{
\setlength{\textfloatsep}{0pt}
\begin{algorithm}[tb]
\scriptsize
\caption{\textbf{\small{Feature extraction}}}
\begin{algorithmic}[1]
\raggedright
\Require event e, K generations for VTCT, D length vector for VTCT, first W words in FCR
\Ensure  FCR(e), VTCD(e), and TXCT(e)
\State FCR(e)=0 \\
\Procedure{Update FCR}{$e$}:
\If {event e is selected as action}
\State FCR(e)+=1
\EndIf
\EndProcedure
\Procedure{Extract FCR}{$e$}:
\State return FCR(e)
\EndProcedure
\Procedure{Extract VTCD}{$e$}:
\State FCR=[][0,0,0,0,0...] /*D length vector with all 0*/
\For {m=0 to K-1:}
\State eventList=BFS(e,m) 
\State /*Use Breadth-first search to extract event List on m-th generation of e*/
\For {childE in eventList}
\State FCR[m][FCR(childE)]+=1
\EndFor
\EndFor
\State return FCR
\EndProcedure
\Procedure{Extract TXCT}{$e$}:
\State TXCT=[]
\State wordList=extractWord(e, W) /*Extract first W words on the text of e*/
\For {word in wordList:}
\State wordVec=word2Vec(word) /*Use word2vec to transfer a word to a vector*/
\State TXCT.append(wordVec)
\State return TXCT
\EndFor
\EndProcedure
\end{algorithmic}
\label{alg:feature}
\end{algorithm}
}



%

\begin{figure}[t]
\vspace*{-5pt}
\centering
\scalebox{0.8}{
\includegraphics[scale=0.45]{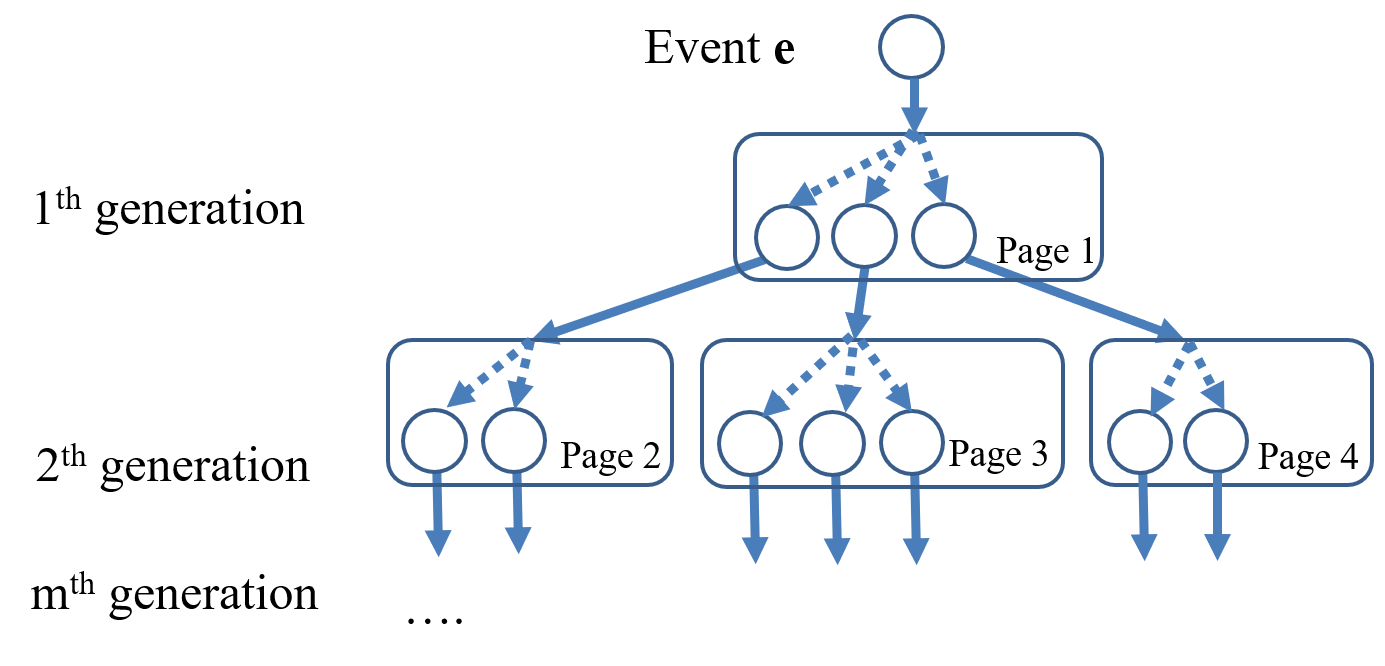}
}
\vspace*{-10pt}
\caption{FCD extraction}
\label{fig:FCD}
\end{figure}

\subB{Execution frequency of children events (FCD).}
A human tester often makes a decision
on which event to execute not only based on the events of  current page,
but also the succeeding pages (children pages)
after executing an event. 
For example, if executing an event $e$ brings the app
into a page containing many unexecuted events, 
$e$ is likely to be selected over the other events
that bring the app into a page containing all 
events already being executed. 
Toward this end, \Name{} uses the execution
frequencies of events on children pages as a feature
for each event of the current page. 
The $m^{th}$ ($m>0$) generation of children pages are defined as the succeeding pages with distance $m$ from the current event 
in the event flow graph.
As the example shown in Fig. \ref{fig:FCD} , the first generation of children pages contains
only one page (i.e., $Page_1$), which results from executing the
selected event (by the agent) on the current page. 
The second generation of children pages includes
pages resulting from executing all events on the page
($Page_1$).  \Name{}'s FCD feature encodes the execution frequencies of events in the $K^{th}$ generation of children pages.
By default $K$ = 3, as we observed that
if $K$ $>$ 3, it almost does not affect the selection of an event.



\Name{} employs a fixed length feature vector 
for each generation of children pages.
Intuitively,
each element in the vector represents an event
on the current generation and the value of the element
records the execution frequency of the event. 
However, a generation of children pages may 
contain a number of events; encoding them all into a feature vector may lead to an unbearable size of vectors 
for a neural network to train.
%
Instead, \Name{} encodes each element in the vector as the number of events
in the current generation that were executed $N$ times, where 
$N$ is equal to the index of the vector. 
By default, the length of
the vector is set to 10. 
%
%
For example, if there is a total of 20 events in the current generation $m$, where 12 events were never executed and 8 events were executed twice, the feature vector 
of $m$ will be $[12,0,8,0,0,0,0,0,0,0]$.   Since $K$ 
is considered by \Name{}, it will generate $K$  vectors for the FCD feature. 


\ignore{
based on the next page (child page).
If exercising current event $e$ can trigger a new page involving events that can achieve higher code coverage, 
$e$ should be selected to exercise.
The EFG constructed by \Name{} is used to extract the feature of children pages 
triggered by current event.  
A straightforward way is considering the the visited times of all events in the children pages
in the feature vector.  
\commentty{How many levels do you consider?}
For example, the current event  triggers a page containing N events, which in turn  triggers N pages with each
containing M events. Therefore, N (M+1) events will be added to the feature vector. 
However, it is impractical for a neural network to train 
a large number of features in a reasonable amount of time. 


To address this problem, 
\Name{} only extracts visited frequency as feature to represent the descendants in the current version. 
It will save space and relief the training burden of neural networks. \Name{} will read the table to find the child page of a current event. 
Then it counts the number of the events which are visited for 0 times, 1 times to 10 times. If an event is visited more than 10 times, it will be considered as 10 times. Then \Name{} would build a vector to record the count. The first item of the vector is the number of 0 visited frequency events. Similar the second item is the number of 1 visited frequency events. Because we set 10 as the up boundary, so the vector's length is 10. \Name{} will consider 3 generation descendants visited frequency as the feature. Every event in the first generation will connect a new page which involves new events as the second generation feature. We use BFS to explore the second generation and third generation. On each generation, \Name{} record the count number on the 10 length vector. So totally, \Name{} would build 3 vectors as features to represent nei-1(generation-1), nei-2, and nei-3's events. Notice that, when \Name{} explores nei-2 and nei-3's events in the graph, some events may be reduplicated to visit. Every event involved its merged event as introduced in last subsection, \Name{} just visit and count one times to avoid the reduplication.
\commentty{Do not understand...}
}

\subB{Textual content of events (TXC).}
The textual content of  events may help 
the agent make the decision on which event to select.  
In the example of Fig.~\ref{fig:tree},
the meaning of ``OK" event  suggests that it has a higher chance to 
bring the app to a new page than the other events. 
To obtain the TXC feature for each event, \Name{}
first employs Word2Vec~\cite{levy2015improving} to convert each word
in the event to a vector with a fixed length $L$. 
The Word2Vec  model is trained from a public dataset text8 containing 
16 million words and is provided along with the source code of Word2Vec~\cite{word2vecgit}.

The text on an event widget can have more than one words, such as
``status bar shortcut" in Fig. \ref{fig:graph}. 
Therefore, we use the first $N$ words to build
the TXC feature. If the number of words 
on an widget is smaller than $N$,  \Name{}
fills the vectors with all zeros to pad the words. 
As a result, the  TXC feature is encoded as 
a matrix $L$ * $N$, where $L$=400 and $N=6$ by default.
According to Pennington et al. \cite{pennington2014glove}, 
$L>$300 has stable performance. 
We tried \{300, 400\} in our experiment and found no different results.
We use L=400 because a larger size can better represent the text. 
We set N to be 6 because the word number on events is rarely up to 6 based on our observation.

\yuc{Note that all configuration parameters (e.g., $K$, $L$, and $N$) can be adjusted to be larger values by users to achieve potentially better coverage, but
at the cost of substantial amount of time in
training and testing. }

\ignore{

 \Name{} employs Natural Language
Processing (NLP) to encode event texts into a feature vector.
It uses Word2Vec~\cite{levy2015improving} to convert the word to a vector with length of 400. 
The Word2Vec  model is trained from a public dataset text8 containing 
16 million words and is provided along with the source code of Word2Vec~\cite{word2vecgit}. 
Sometimes, an event may have hundreds words in the text. \commentty{why?}
To improve the efficiency and accuracy of training, \Name{} only uses first $N$ words
in the text of event. If the words on an event is shorter than $N$, \Name{} sets a padding as all 0 
vector to represent them (By default $N$=6). After that, text on every event on current page is converted to an $N$*400 matrix.
\commentty{How is 400 chosen?}. The matrix is used as the feature input to the neural network.  
\commentty{How is word to vec used?}
}

\subsubsection{Compacted Event Flow Graph}
\label{subsubsec:efg}

\begin{figure}[t]
\vspace*{-5pt}
\centering
\scalebox{0.8}{
\includegraphics[scale=0.45]{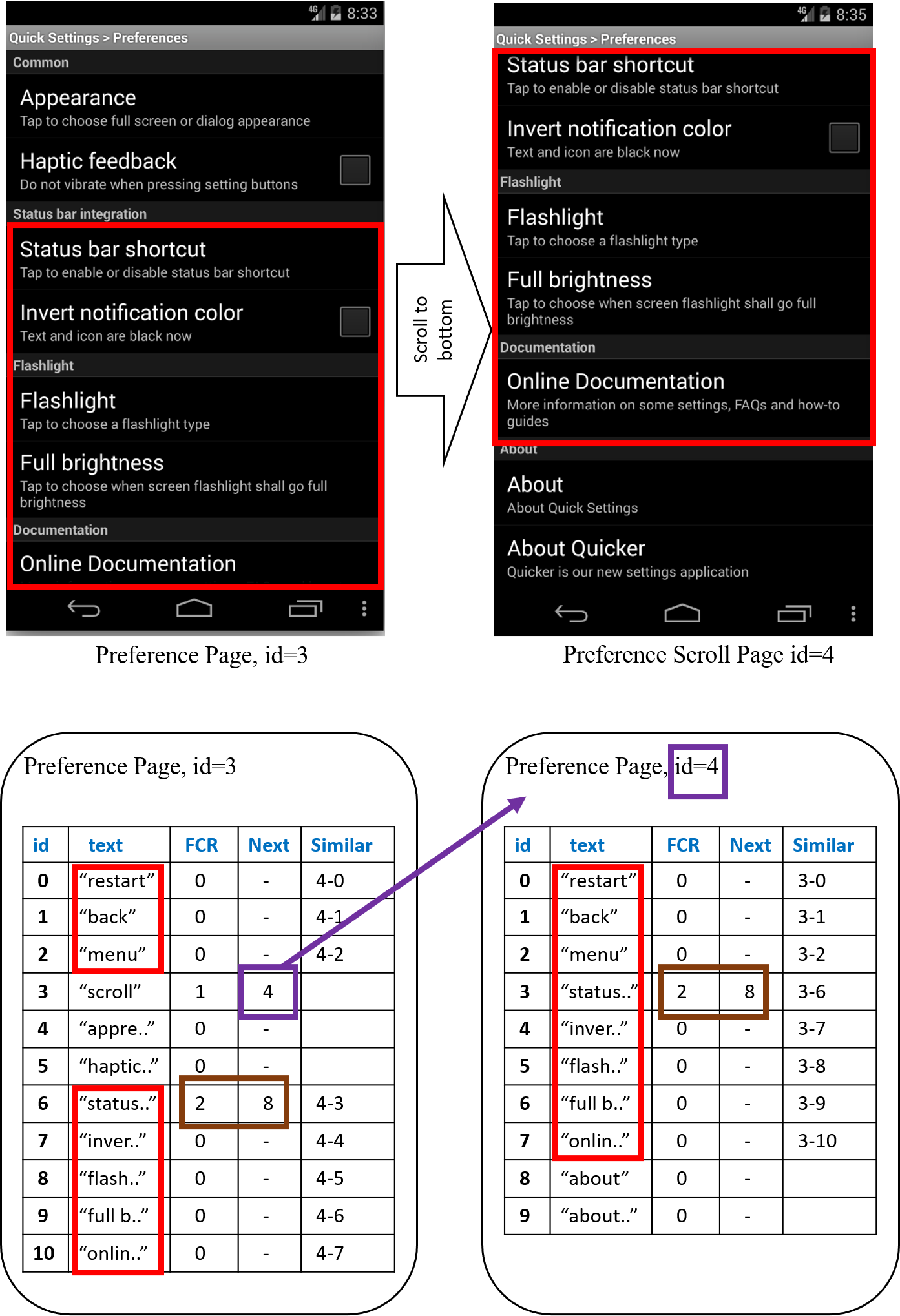}
}
\vspace*{-5pt}
\caption{Event Flow Graph Example }
\label{fig:graph}
\end{figure}

\Name{} uses an event flow graph (EFG) to obtain features. 
The graph is represented by $G$ = ($V$, $E$). 
The set of vertices, $V$, represent events using
their unique IDs. The
set of edges, $E$, represent event transitions (i.e., from the selected event $e$ to all events in the next page) triggered by executing $e$.
Each vertex records the three types of features described in Section~\ref{subsubsec:featureType}.
Fig.~\ref{fig:graph} shows an example, where each page is associated
with a list of vertices and their
features. The last label (``Similar")
will be discussed later in the section. 

\Name{}'s event flow graph is compacted in order to accommodate DQN. 
This is also the main difference from traditional EFGs~\cite{amalfitano2012using} . 
In traditional EFG,  whenever a new page  is encountered, it
will be added as a vertex to the EFG. 
However, if we use the vertices as states in DQN, it could cause the states
to be huge or even unbounded~\cite{baek2016automated, choi2013guided, memon2003gui, pradel2014eventbreak, su2017guided}. 
Also, when encountering a similar page  with a minor difference
from an earlier page, if \Name{} treats it as a new state, it could 
generate unbounded fake new events and thus waste exploration time.
For example, in Fig. ~\ref{fig:graph}, there are many same events on the pages triggered
by the ``scroll" event. If we consider all of the events for each scroll triggered
page, \Name{} will waste time to execute the same events again and again
without code coverage increase. 
%
Therefore, such similar
pages should be combined to avoid this problem. 

There have been techniques on compacting EFG for efficient exploration. 
For example, Stoat~\cite{su2017guided} ignores the details of ListView events 
by categorizing it into ``empty" and ``non-empty", so it can merge two 
similar pages with only the ListView events are different into
the same state.  However, it may lose important information since some of 
the events triggered by the items under the ListView may be critical. 
%
%
In contrast, \Name{} compacts the EFG by merging vertices instead of 
pages. Specifically, whenever a new page $P'$ is encountered,  
\Name{} retrieves the pages
with the same Android Activity ID ~\cite{Activity}  using UI Automator ~\cite{uiautomator} from the existing
EFG.  
It then compares the text of each vertex in $P'$ with the vertices of each of
the retrieved pages.  
If the texts are the same, the two events are merged
into a single vertex.  The ``Similar" tag in Fig.~\ref{fig:graph} records the information
of the merged vertex (i.e., page id---event id). 
As such, when an event $e$ is executed,  the feature vectors/matrices 
of both $e$'s vertex and its merged vertex are updated. 
Therefore, the same events will get an equal chance to be executed. 

\ignore{
whenever a new page $P'$ is triggered, \Name{} compares 1)
the activity names of $P'$ and the previous page $P$, and 2)  the 
the text of each vertex in $P'$  and $P$. \commentty{what is an activity name?}

can retrieve accurate information of pages by merging 
the vertices at the event level rather than at the page level. Specifically, 
if \Name{} encounters a similar page but not the exactly same page with an existing table, 
it will build a new table but record and merge the same event occurs on two pages.
\commentty{How to determine if the pages are similar.}
 For example, in Fig.~\ref{?}, there are five listViews items  ("Status bar shortcut", "Invert notification color", "Flashlight", "Full brightness", and "Online Documentation") 
 on the pages are the same texts on the two different pages as "id=3","id=4".  
 Because of the activity name of page is same and some text on the page is same, \Name{} considers the items with 
 same texts are the same events and extends to the system default events "restart", "back", "menu" be same events too. 
 \commentty{what does that mean?}
 Then \Name{} fill the similar column of each table to record the merge information as pageId-eventId. 
 For example, the event "Status bar shortcut" id=6 of page add a similar event 4-3, which means the event with id=3 on page id=4 
 is the same event with the current event. After that, when \Name{} updates one event's visited frequency and next, it will update 
 all its similar event at the same time. As shown in Fig. 4, the red square's highlight events have same Freq and next value. Because the similar pages are not merged into a state, \Name{} can still read from the graph about the next page information from an event. Although the event level merge is not our main contribution, \Name{} is the first work to do that as far as we know.
}

\begin{figure*}[t]
 \centering
 \includegraphics[scale=0.16]{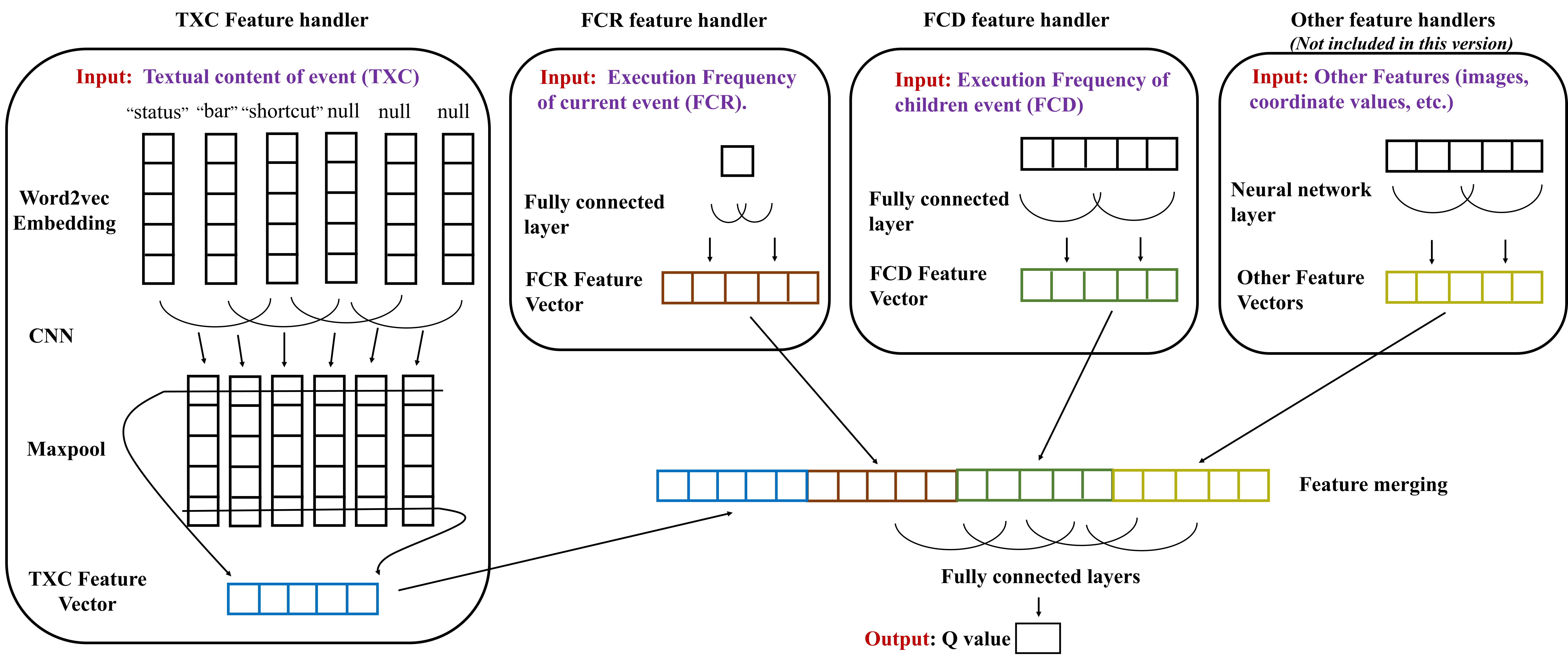}
  \caption{Deep Q-Network Model}
  \label{fig:model}
  \vspace*{-5pt}
\end{figure*}

\subsection{\Name{}'s Deep Q-Networks}

\Name{}'s DQN agent examines the current state of the app and performs an action, i.e., selects the event with the largest Q value. 

\subsubsection{DNN model}

One of the key components in \Name{} is a 
Deep Neural Network (DNN) model,
which uses equation (1) to compute the Q value. 
The DNN takes the features of the event from last 
action $a_t$ as input and outputs 
the Q value, which is equal to  $r_t+\gamma*\underset {a} \max Q(s_{t+1},a)$. 
%
Comparing to regular RL using a Q-table, DNN is 
able to calculate Q values from any dimensions of features for each event. 
The DNN model involves feature handling and 
feature merging.  The feature handling component
employs a neural network algorithm to 
process each individual feature
into a specific modality, represented
by a one-dimension vector. 
The feature merging component
combines the vectors of all features and passes
them into the final fully connected layer~\cite{wan2013regularization} that is used to determine
the event's Q value. 
The purpose of the feature handling
process is to help DNN make  an
accurate prediction.


\subB{Feature handling.}
\Name{} employs specific DNN sub-models
to process individual features according to their types. 
For example, as shown in Fig.~\ref{fig:model},
the TXC  feature is processed
by a convolutional neural network (CNN)
based on natural language processing (NLP)
and maxpool~\cite{kim2014convolutional}. 
The FCR feature is handled by 
a fully connected layer~\cite{wan2013regularization},
which is a common and efficient layer in neural network to 
process vectors. Other algorithms~\cite{justus2018predicting} could also be 
used but may involve additional cost. 


\ignore{
A feature handler transforms a feature 
value into a vector, 
For example, in Fig.~\ref{fig:model}, embedding vectors of text "status bar shortcut" are input into the 
CNN and the maxpool will output a vector as the pre-process result~\cite{kim2014convolutional}. 
The output is a vector representing the text feature. 
\Name{} can be easily extended by writing new handlers to process other features.
For example, users can take image as another type of feature by leveraging
existing image classification models~\cite{?}.
The DNN model accepts the formats of single value, vector, and matrix. 
By default, single value and vector are handled by a fully connected layer
and matrix is handled by 1-D CNN ~\cite{Conv1D}.

\yuc{
\Name{} uses \textit{feature handlers} to process features. For example, in Fig.~\ref{fig:model}, after feature generation, the event(button) "status bar shortcut" on Fig. \ref{fig:graph} are represented by multiple features as FCR, VTCD, and TXCT. For each feature, \Name{} use one unique and dedicated feature handlers to process each feature. For example, we use a classical natural language processing CNN and maxpool based model as the feature handler 1 to process the TXCT. The output is a vector as the pre-process result~\cite{kim2014convolutional} which represents the text feature. }
}

\Name{}'s DNN model can be easily extended to process other features.
For example, users can take image as another type of feature by leveraging
existing image classification algorithms~\cite{wang2016cnn}. 
The DNN model accepts the formats of single value, vector, and matrix. 
By default, single value and vector are handled by a fully connected layer
and matrix is handled by 1-D CNN ~\cite{Conv1D}. 


\subB{Feature merging.}
Since the DNN model predicts a Q value based on all features,
it combines the vectors generated by the feature handling components
into a large one-dimensional vector. 
\yuc{All input features to the DNN share the unique loss function described in Section 2.2.2.
We leverage Keras deep learning architecture, which
is capable of accepting mixed multiple inputs~\cite{gao2018hyperspectral}.}
It then utilizes a three-fully connected
layers~\cite{Dense} to process the large vector and predicts a Q value. 
 The last layer with a linear activation to output just one value to represent the Q value. 
 We use a stochastic gradient descent algorithm, Adam~\cite{adam} to optimize the model with a
 learning rate 0.0001. The Adam optimization algorithm is an extension of stochastic gradient 
 descent, which has been widely adopted for deep learning applications in computer vision and natural language processing~\cite{ali2019adam,gregor2015draw}. 
 

\subsubsection{Action Selection}
\label{subsubsec:event}


\Name{}'s DQN agent performs an action by selecting an event to execute from
the current page. For editable events (e.g. text fields), \Name{} automatically 
generates a random string with digit, char, and punctuation. 
\Name{} maintains a Q value for each event.
%
At the first $T$ iterations, \Name{}  executes
random events to construct the DQN model.
Randomness is necessary
for an agent navigating through a stochastic maze to learn the
optimal policy ~\cite{tijsma2016comparing}. 
%
%
By default, $T$=20. 
After that,  \Name{}  starts to use DNN to select an event.


To determine which event to execute on the current page, 
\Name{} uses the $\varepsilon$-greedy policy~\cite{tokic2011value}, a widely adopted policy in reinforcement learning, to select the next event. \Name{}  selects the event with the highest Q value 
with probability $1-\varepsilon$
and a random event with probability $\varepsilon$. 
The Q value is computed by the neural network model (Section~\ref{subsubsec:dql}). 
The value of $\varepsilon$ can be adjusted by users.  By default $\varepsilon$ = 0.2,
which is the same as the value used in Q-testing ~\cite{pan2020reinforcement}. 
In order to amplify the chance of
bug detection, \Name{} also generates system-level events
at every 10 iterations, such as screen rotation, volume control,
and phone calls~\cite{su2017guided}. 
\ignore{
\begin{equation}
getActionEvent(s)=
\left\{
             \begin{array}{lr}
             random \ e \in s, & first \ 20 \ iterations  \\
             arg \ max_e Q(s, e), & 1-\varepsilon\\
             random \ e \in s, & \varepsilon 
             \end{array}
\right.
\end{equation}
}

\subsubsection{Reward Function}

In reinforcement learning, a reward is used to
guide training and testing.  \Name{}'s reward function is based on code coverage and bug detection.
Specifically, \Name{}  sets the reward to a positive number 
(r=5) when the code coverage increases or revealing a unique crash
and a negative number (r=-2) when the coverage does not change. 
To determine if a unique crash is revealed, we
resort to the error message. 
In the log of an app execution, 
each crash is associated an accurate error message.
If the same error message has happened before,
we ignore it. 
The absolute value of positive reward is larger than that of the 
negative reward is because the intention is letting the machine favor higher
coverage or detecting bugs. The reward values
are configurable, however, these default values work well on
different types of apps, as shown in the experiments.



Other reward functions, such as measuring the app page state changes
may also be used~\cite{pan2020reinforcement, koroglu2018qbe, degott2019learning, adamo2018reinforcement, vuong2018reinforcement}. 
For example, Q-testing~\cite{pan2020reinforcement} calculates the difference 
between the current page state and the states of recorded pages.

\ignore{
\yuc{While triggering unvisited page states may have
a positive relationship with higher code coverage,
it is not a direct measurement. It is very likely that new unvisited page states do not lead to code coverage increase. In Fig. ~\ref{fig:graph},  the scroll event can reach a new page state but it does not increase code coverage. On the other hand, an already visited page state, when being visited again,
could lead to coverage increase due to a different event flow path.
For example, in a music play app, every time the `play" is clicked
to play a different song to visit the same play page, the code coverage may increase.
Therefore, \Name{} uses code coverage directly to calculate reward while encoding
the indirect measures (e.g., execution frequency) as a type of feature.}
}

\ignore{Toward this end, \Name{} designs a new reward function by directly measuring code coverage.
Specifically, \Name{}  sets the reward to a positive number when the code coverage increases 
and a negative number when it decreases. The reason is  
because by reading complex features, \Name{}  can let the agent  learn complex behaviors about 
how to increase code coverage.  \commentty{Need to elaborate the challenge. And evaluation?}
}

\section{Evaluation}
\label{sec:evaluation}

To evaluate \Name{}, we consider three research questions:


\vspace*{3pt}
\noindent
{\bf RQ1:}
How does  \Name{} compare with the state-of-the-art Android testing tools in terms of code coverage?

\vspace*{3pt}
\noindent
{\bf RQ2:} 
How does  \Name{} compare with the state-of-the-art Android testing tools in terms of bug detection?

\vspace*{3pt}
\noindent
{\bf RQ3:}
Can \Name{} understand the features and learn a correct model?

\subsection{Datasets}


We need to prepare datasets for evaluating our approach. 
Since Sapienz~\cite{mao2016sapienz} and 
Stoat~\cite{su2017guided} are two of the baseline 
tools that contain publicly available 
datasets to compare with, we used the 
datasets from their papers~\cite{choudhary2015automated}.  
The datasets contains a total of 68 apps that falls into 18 domains~\cite{EvaluationDataSet}.
We removed four apps because
they crash right after launch on our Intel Atom(x86) emulator.
The executable lines of code in the apps range from 109 to 22,208, indicating
that they represent apps with different levels of complexity.

\subsection{Implementation}

We conducted our experiment on a four-core 3.60 GHZ CPU 
(with no GPU acceleration) physical x86 machine running with Ubuntu 16.04. With this machine, \Name{} can send an event in 1.43 seconds on average.
The dynamic exploration component
is implemented on UI Automator~\cite{uiautomator}. 
The system events issued during testing are obtained from 
Androguard~\cite{androguard}. \Name{}  uses 
Emma~\cite{emma} to obtain statement coverage. 
Keras~\cite{Keras} is used to build and run the deep neural network.
The DQN agent is implemented by ourselves using  Python.

\subsection{Study Operation}

We performed a two-fold cross validation by randomly dividing the whole 64 apps data set into two sets with each set containing 32 apps. 
It took 128 hours finish the experiment. 
An added fold would take about three more days. 
%
%
Specifically, 
the process 
randomly picked 32 apps as the training set and 
used the remaining 32 apps as the testing set.  
\yuc{In the next fold, the first 32 apps are used in the testing set and the other 32 apps 
in the training set. Therefore, each app has a chance to be
involved in both the training set and the testing set.} 
The testing time of \Name{} is set to one hour \yuc{for each app}, which
is consistent with many other Android app testing tools~\cite{mao2016sapienz, li2019humanoid, pan2020reinforcement, vuong2018reinforcement}.
\ignore{The testing time is  further divided into three phases 
with each phase 20 minutes.
For each phase, \Name{} used the previously learned model
as the initial model to test the app and update the model. \yuc{Each phase is a process that \Name{} adapt or update the before trained model to better fit the app under test.} 
\yuc{Remove:By dividing the testing into multiple phases, it could avoid \Name{} from
repeatedly exploring the same events.} \yuc{Add (We can remove it if you think it is not good): By dividing the testing into multiple phases, it could provide chances for \Name{} to learn about how to test new apps in different phases. It can avoid \Name{} to get stuck or do wrong behaviors at the first phase e.g. get stuck in a loop or repeatedly restart the app. For example, we find \Name{} considers it has no interest on any of event on the current page sometimes, then it keep clicking restart or a broken button. This thing mostly happens at the first phase with a bad initial exploration in first 20 random event selection. At that time, \Name{} is not familiar with the new AUT at the first phase. We do not use heuristic to restrict \Name{}'s behavior. We only give \Name{} chances to know how to test a new app. If at the first phase \Name{} gets stuck at somewhere, \Name could summarize the before experience and learn and test better at the next phase.}
\commentty{Need clarities.}
}
Given that RL has the tendency to exhibit high variability across runs, we repeat the testing for three times.  The results are averaged over the 3 executions of each tool.


\subsection{Comparison with Existing Tools}

We compared \Name{} with five existing Android
testing tools: Monkey~\cite{monkey} , Stoat~\cite{su2017guided}, Sapienz~\cite{mao2016sapienz}, and QBE~\cite{koroglu2018qbe}, Q-testing~\cite{pan2020reinforcement}. 
Monkey is a random testing tool. 
\yuc{Stoat employs model-based testing and pre-defined heuristic methods to explore 
the less unexecuted actions.} Sapienz employs evolutionary testing. 
Both QBE and Q-testing use
traditional Q-learning and QBE can also train and test on different sets of apps. 
The details of the tools are
discussed in Section~\ref{sec:related}. 

\ignore{
it can transfer the knowledge learned from 
the training set to new apps without manual labeling work
(Section~\ref{subsec:limitations}). \yuc{We do not choose Q testing ~\cite{pan2020reinforcement} to compare with because Q testing can not transfer knowledge among apps. Q testing's main focus is the reward design which is not conflicted with \Name{}. \Name{} can utilize Q testing's reward design to improve the performance.}
\commentty{we do not want to emphasize transfer. re-implement QBE.}
}

The testing time for all tools is set to one hour. Like \Name{}, we repeated the
experiment three times. 
We followed previous work~\cite{choudhary2015automated, pan2020reinforcement} to set  200 milliseconds delay between events for Monkey to avoid abnormal behaviors. 
Stoat~\cite{su2017guided} allocates one hour for model construction phase
and two hours for Gibbs sampling phase.  To use Stoat in our one-hour testing,
we followed the suggestion proposed in \cite{pan2020reinforcement}, in which
each phase of Stoat is set to 30 minutes for the best performance. 

\yuc{Note that while the training time 
of \Name{} is extensive, we believe 
the cost will be diluted as the number of apps increases. 
If new apps are added to the training set, the existing
model is directly updated without training from scratch.
For the existing non-learning-based tools
(e.g., Monkey, Sapienz), updating pre-defined rules often needs huge professional human effort. 
}

\section{Results and Analysis}
\label{sec:results}

\subsection{RQ1: Code Coverage}

Table ~\ref{tab:coverage} shows the results of 
code coverage obtained from \Name{} and the other
five tools on  64 apps.  For each tool, we compute the
average of the coverage scores collected in 
three test runs. The test results and log files can be found in 
our experiment dataset~\cite{DinoDroid}.

On average, \Name{} achieves 48.9\% line coverage, 
which is 22.8\%, 17.8\% 13.7\%, 16.1\%, and 18.6\% more effective
than Monkey, Q-testing, QBE, Stoat, and Sapienz, respectively. 
Specifically,  \Name{} achieved the highest coverage
in 33 apps, compared to 16 apps in Sapienz, 9 
in Stoat, 10 in Q-testing, 8 in QBE, and 5 in Monkey.  
The results indicate \textit{\Name{} is effective
in achieving high coverage.}

We analyzed the app code covered by different tools. 
Taking the app ``Nectroid"~\cite{nectroid} as an example, 
many events reside in deep levels of EFG, which
require a number of steps to explore. For instance,
two widgets are associated with important functionalities: ``add a new site" and ``delete a site". 
Exercising the first widget requires 7 steps
(launched page $\rightarrow$ menu $\rightarrow$ settings $\rightarrow$ select a site $\rightarrow$ new site $\rightarrow$ fill blankets $\rightarrow$ OK)
and exercising the second widget requires 6 steps (launched page $\rightarrow$ menu $\rightarrow$ settings $\rightarrow$ select a site $\rightarrow$ Nectarine(long click) $\rightarrow$ delete $\rightarrow$ OK).  
Monkey, QBE, and Sapienz failed to cover the two 
particular sequences because they were not able to 
navigate the app to the deep levels. Q-testing failed to reach ``delete a site". 
It finished ``add a new site", but was not able to fill in any text information of the new site.
Stoat was very close to reach ``add a site", but
at the last step,  it selected ``Cancel" instead of ``OK".  
On the other hand, 
\Name{} exercised the ``OK" widget in 90\% of pages
that contain ``OK", ``Cancel", and a few other functional
widgets during the learning process. 
We conjecture that this is because \Name{} is able 
to  learn that clicking ``OK" widget is more likely to increase coverage. 


\ignore{
understand
the content of widgets from learning the previous apps. \yuc{ We find the \Name{} chooses to select functional button "OK" rather than "Cancel" then achieves the functionalities in this example.  We summarize all of cases of \Name{} for "Cancel" and functional buttons in 64 APPs testing trace. The functional buttons may be "OK", "Delete", "Login", "Search", and so on. In 233 cases, on the page which involves only "Cancel" and only one functional button. \Name{} has 8.6\% cases to only execute cancel,  25.3\% cases to only execute functional button, 62.2\% cases to execute both of buttons, 3.8\% cases to execute a system widget rather than these two buttons. It means \Name{} only has 12.4\% to miss the functional button and executes others in a page.}
}




\ignore{
As shown in Fig. ~\ref{fig:coverageBox}) on average, \Name{} gets 49\% line coverage more than 
Monkey (40.3\%), Stoat(42.8\%), Sapienz (42.6\%). Additionally, \Name{} achieves the highest code 
coverage on 33 apps which is more than Monkey (4), Stoat (13) and, Sapienz (20). 
Specifically, \Name{} achieves 32 highest code coverage in apps whereas Sapienz 20 , Stoat 12, QBE 3, Monkey 3. 
For the first 34 apps with ELOC from 109 to 1242, \Name{} wins 14 Apps. 
For the second 34 apps with ELOC from 1245 to 22208, \Name{} wins 18 Apps. 
\Name() wins more on the app with higher ELOC.

\commentty{From existing paper? We issued less number of events but better than them...}
Monkey, QBE, Stoat, Sapienz, and \Name{} averagely generate approximately 14235, 1456, 2013, 
12992 and 3533 events. As for Monkey, there is no need for it to analyze
widget hierarchy before generating events. Besides, it can communicate directly with several essential Android services. 
These features enable it to generate events with high speed. Even if Monkey can
generate many more events than others, most of them are redundant and may make 
no effect on the applications. Sapienz comes second in generating events as it is built upon Monkey. 
It is encouraging to
see the number of executed events by \Name{} is close to that of stoat's and QBE. 
All of these three tools need to extract the event information from pages of APP and select one to execute. 
}

\subsection{RQ2: Bug Detection}
Table~\ref{tab:coverage} shows the number of unique bugs  detected
by the six tools (on the left of "/").  We reported the number of
unique bugs detected in all three repeated runs. 
Like existing tools~\cite{su2017guided,mao2016sapienz,pan2020reinforcement}, we consider a bug to be a crash or an exception.
The results showed that \Name{}  detected the largest number of bugs (269) compared to Monkey (73), Q-testing (54), QBE (59), Stoat (189), and Sapienz (63)).
Specifically, \Name{} detected most bugs in 38 apps, which is more effective
than Monkey (5), Q-testing (0), QBE (3), Stoat (20), and Sapnize (2). 

\Name{} and Stoat detected a significantly larger number of bugs because they both use androguard~\cite{androguard}
to issue system-level events to amplify the chance of bug detection (Section~\ref{subsubsec:event}). 
When disabling system-level events, \Name{} still
detected the largest number of bugs (130), 
compared to Monkey (72), Q-testing (47), QBE (52), Stoat (78), and Sapienz (59). 
The numbers are reported in the right side of ``/" in Table~\ref{tab:coverage}. 
Specifically, \Name{} detected most number of bugs in 26 apps, which is more effective
than Monkey (13), Q-testing (4), QBE (6), Stoat (11), and Sapnize (9).
The above results  suggest that \textit{\Name{} is effective
in detecting bugs.}

\ignore{\yuc{Adding system-level events sometimes causes unrealistic crashes, for example, sending system-level events to start an activity that requires specific data, without required data, the app would crash with exceptions such as ClassNotFoundException or NullPointerException. The crashes caused by sending such system-level events seem to be unrealistic and unlikely to be triggered by normal users.}
Therefore, for a fair comparison with Monkey, Sapienz, and QBE, 
we removed the bugs relevant to these system
level events from the five tools \yuc{by ignoring crashes logs with 4 statements ~\cite{DinoDroid}}.

The results (on the right of ``/")still showed that \Name{} detected the largest number of bugs (130), 
compared to Monkey (72), QBE (52), Stoat(78), and Sapienz(59). 
}

\ignore{The faults are defined as the Fatal exceptions on the App under test. \Name{} reveal the highest number of faults as 87 comparing to Monkey(25), Stoat(62), Sapienz(21). By analyzing the fault source, we find the reason that stoat and \Name{} achieve very big advantage than others because stoat and \Name{} uses androguard~\cite{androguard} to analyze  the events that the apps are particularly interested in, which are usually declared by the tags <intent-filter> and <service> in their AndroidManifest.xml files. The biggest difference between stoat, \Name{} with others are from these kind of events triggered crash. To just consider the event exploration triggered crash, we do not count these crashes from system-level events. These crashes have particular crash log as below. After we remove these crashes from all of these tools, \Name{} still reveals the highest number of faults as 41 comparing to Monkey (24), Stoat(26), and Sapienz(19). \commentty{What are the items below? The tool can issue system events, causing crash...}

(1) android.app.ActivityThread.handleReceiver

(2) at android.app.ActivityThread.performLaunchActivity

(3) at android.app.ActivityThread.handleServiceArgs

(4) android.app.ActivityThread.performResumeActivity

For 64 apps, \Name() triggers the argest number of  faults in 23 apps whereas Sapnize 7, Stoat 13, QBE 6, Monkey 13 without the system intent crash. \Name() triggers the argest number of faults in 42 apps whereas Sapnize 5, Stoat 28, QBE 3, Monkey 8 without the system intent crash.
}

Figure~\ref{fig:crash} shows the pairwise comparison of bug detection results
between tools (with system-level events) in a single run. \yuc{The black bars indicate the number of bugs detected in common and the grey bars indicate the number of bugs detected uniquely in each tool. }
%
%
For example,  Stoat and \Name{} detected
50 bugs in common. However,  \Name{} detected 37 bugs not detected by Stoat and
Stoat detected 12 bugs not detected by \Name{}. 



\ignore{

\begin{table}[t]
\centering
\scriptsize
\setlength{\tabcolsep}{3pt}
\caption{\label{tab:coverage} Testing Result for Comparison}
\scalebox{0.7}{
\begin{tabular}{|l|c|c c c c c|c c c c c|} \hline
  \#{\em APP.}
& {\em \# LOC}  
& \multicolumn{5}{c|}{\em Code Coverage}
& \multicolumn{5}{c|}{\em \# Fault Triggered}
\\
&
& {\em Mon.} &{\em QBE.} &{\ \em St.\ } & {\em Sap.}  & {\em RD.}  
& {\em Mon.} &{\em QBE.} &{\ \em St.\ } & {\em Sap.}  & {\em RD.}  
\\ \hline
soundboard & 109  &31  & 42 & 42 &\cellcolor{gray!30} 63 & 42 & 0/0 & 0/0 & 0/0 & 0/0 & 0/0 \\ 
gestures & 121  & 26 & 32 & 32 &\cellcolor{gray!30} 52 & 32  & 0/0 & 0/0 & 0/0 & 0/0 & 0/0 \\ 
fileexplorer & 126  & 31 &\cellcolor{gray!30} 40 &\cellcolor{gray!30} 40 & 31 &\cellcolor{gray!30} 40 & 0/0 & 0/0 & 0/0 & 0/0 & 0/0  \\ 
adsdroid & 236  & 8 & 29 & 31 &\cellcolor{gray!30} 33 & 29  &\cellcolor{gray!30} 1/1 & 0/0 & 0/1 &\cellcolor{gray!30} 1/1 & 0/1 \\ 
MunchLife & 254 & 66 & 66 & 66 &\cellcolor{gray!30} 71 & 69  & 0/0 & 0/0 & 0/0 & 0/0 & 0/0 \\ 
Amazed & 340   &66  & 69 & 58 &\cellcolor{gray!30} 74 & 73  &\cellcolor{gray!30} 2/2 & 0/0 & 0/0 & 1/1 & 1/1 \\ 
battery & 342  & 72 & 70 & 69 & 46 &\cellcolor{gray!30} 74  & 0/0 & 0/0 & 0/0 & 0/0 &\cellcolor{gray!30} 1/1 \\ 
manpages & 385 & 58 & 63 & 56 &\cellcolor{gray!30} 69 & 63  & 0/0 & 0/0 & 0/1 & 0/0 & 0/1 \\ 
RandomMusicPlayer & 400 & 53 & 54 &\cellcolor{gray!30} 74 & 57 & 60  & 0/0 & 0/1 & 0/0 & 0/0 & 0/1 \\ 
AnyCut & 436 & 61 & 62 & 61 &\cellcolor{gray!30} 65 & 62  & 0/0 & 0/0 & 0/0 & 0/0 & 0/0 \\ 
autoanswer & 479 & 11 & 13 &\cellcolor{gray!30} 26 & 18 & 22 & 0/0 & 0/0 & 0/1 & 0/0 & 0/1 \\ 
LNM & 492 & 54 & 55 &\cellcolor{gray!30} 61 & 52 & 57  & 0/0 & 0/0 & 0/2 & 0/0 & 0/1 \\ 
baterydog & 556 & 62 & 61 & 54 &\cellcolor{gray!30} 67 & 62  & 0/0 & 0/0 &\cellcolor{gray!30} 1/1 & 0/0 & 0/0 \\ 
yahtzee & 597 & 8 & 46 & 42 & 39 &\cellcolor{gray!30} 58  & 0/0 &\cellcolor{gray!30} 1/1 & 0/1 & 0/0 &\cellcolor{gray!30} 1/2 \\ 
LolcatBuilder & 646 & 27 & 20 & 25 & 31 &\cellcolor{gray!30} 53 & 0/0 & 0/0 & 0/0 & 0/0 & 0/0 \\ 
CounterdownTimer & 650 & 58 & 76 &\cellcolor{gray!30} 77 & 50 &\cellcolor{gray!30} 77  & 0/0 & 0/0 & 0/0 & 0/0 & 0/0 \\ 
lockpatterngenerator & 669  & 78 & 75 & 69 &\cellcolor{gray!30} 79 &\cellcolor{gray!30} 79  & 0/0 & 0/0 & 0/0 & 0/0 & 0/0 \\ 
whohasmystuff & 729 & 65 & 72 & 68 & 32 &\cellcolor{gray!30} 74  & 0/0 & 0/0 & 0/1 & 0/0 & 0/1 \\ 
Translate & 799 & 46 & 44 & 36 &\cellcolor{gray!30} 49 & 47  & 0/0 & 0/0 & 0/0 & 0/0 & 0/0 \\ 
wikipedia & 809 & 24 & 27 & 27 &\cellcolor{gray!30} 28 & 27  & 0/0 & 0/0 & 0/0 & 0/0 &\cellcolor{gray!30} 1/2 \\ 
DivideAndConquer & 814  &\cellcolor{gray!30} 83 & 81 & 52 & 80 & 58  & 0/0 & 0/0 & 0/0 &\cellcolor{gray!30} 1/1 & 0/0 \\ 
zooborns & 817 & 34 & 34 & 36 &\cellcolor{gray!30} 37 & 35  & 0/0 & 0/0 & 0/1 & 0/0 & 0/1 \\ 
multismssender & 828 & 46 & 46 & 51 & 60 &\cellcolor{gray!30} 68  & 0/0 & 0/0 & 0/1 & 0/0 & 0/1 \\ 
Mirrored & 862 & 44 & 45 & 45 &\cellcolor{gray!30} 46 & 45  & 0/0 & 0/0 & 0/1 & 0/0 & 0/1 \\ 
myLock & 885  & 26 & 27 &\cellcolor{gray!30} 44 & 30 & 41  & 0/0 & 0/0 &\cellcolor{gray!30} 1/2 & 0/0 & \cellcolor{gray!30} 1/2 \\ 
aLogCat & 901 & 66 & 68 & 71 & 43 &\cellcolor{gray!30} 79  & 0/0 & 0/0 & 0/0 & 0/0 & 0/0 \\ 
aGrep & 928 & 36 & 46 & 45 & 55 &\cellcolor{gray!30} 56  & 0/1 &\cellcolor{gray!30} 2/3 & 1/1 & 1/1 &\cellcolor{gray!30} 2/3 \\ 
dialer2 & 978 & 37 & 39 & 32 & 40 &\cellcolor{gray!30} 46  & 0/0 & 0/0 & 0/1 & 0/0 & 0/1 \\ 
hndroid & 1038 & 9 & 9 & 9 &\cellcolor{gray!30} 10 & 9  &\cellcolor{gray!30} 1/1 & 0/0 &\cellcolor{gray!30} 1/1 &\cellcolor{gray!30} 1/1 &\cellcolor{gray!30} 1/1 \\ 
Bites & 1060 & 33 & 25 & 46 & 41 &\cellcolor{gray!30} 51  & 1/1 & 0/0 &\cellcolor{gray!30} 4/5 & 1/1 &\cellcolor{gray!30} 4/5 \\ 
tippy & 1083 & 82 & 58 & 72 & 82 &\cellcolor{gray!30} 84  & 0/0 & 0/0 & 0/0 & 0/0 & 0/0 \\ 
weight$-$chart & 1116 & 55 & 62 & 46 & 58 &\cellcolor{gray!30} 78  & 0/0 & 0/0 &\cellcolor{gray!30} 1/1 & 0/0 &\cellcolor{gray!30} 1/1 \\ 
importcontacts & 1139 & 40 & 41 & 31 &\cellcolor{gray!30} 42 & 41  & 0/0 & 0/0 & 0/0 & 0/0 & 0/0 \\ 
worldclock & 1242 & 89 &\cellcolor{gray!30} 92 &\cellcolor{gray!30} 92 & 90 & 91 & 0/0 & 0/0 & 0/1 & 0/0 & 0/1 \\ 
blokish & 1245 & 41 &\cellcolor{gray!30} 51 & 36 & 44 & 50 &\cellcolor{gray!30} 1/1 &\cellcolor{gray!30} 1/1 & 0/0 & 0/0 &\cellcolor{gray!30} 1/1 \\ 
aka & 1307 & 56 & 80 & 79 &\cellcolor{gray!30} 81 & 64  & 1/1 & 1/1 & 0/0 & 2/2 &\cellcolor{gray!30} 3/3 \\ 
Photostream & 1375 & 24 & 14 & 24 &\cellcolor{gray!30} 28 & 21  & 1/1 & 1/1 &\cellcolor{gray!30} 2/3 & 1/1 & 1/2 \\ 
dalvik$-$explorer & 1375 & 43 & 70 & 70 & 69 &\cellcolor{gray!30} 71  &\cellcolor{gray!30} 1/1 &\cellcolor{gray!30} 1/2 & 0/1 &\cellcolor{gray!30} 1/2 &\cellcolor{gray!30} 1/2 \\ 
tomdroid & 1519 & 51 & 50 &\cellcolor{gray!30} 53 & 51 &\cellcolor{gray!30} 53  & 0/0 & 0/0 & 0/0 & 0/0 & \cellcolor{gray!30} 1/1 \\ 
PasswordMaker & 1535 &\cellcolor{gray!30} 62 & 55 & 55 & 37 & 56  & 0/0 &\cellcolor{gray!30} 3/3 &\cellcolor{gray!30} 3/4 & 1/1 & 2/3 \\ 
frozenbubble & 1706 &\cellcolor{gray!30} 86 & 59 & 55 & 81 & 69  & 0/0 & 0/0 & 0/0 & 0/0 & 0/0 \\ 
aarddict & 2197 & 13 & 13 &\cellcolor{gray!30} 31 & 14 & 18 & 0/0 & 0/0 &\cellcolor{gray!30} 2/2 & 0/0 & 0/0 \\ 
swiftp & 2214  & 13 & 12 & 13 &\cellcolor{gray!30} 14 & 13  & 0/0 & 0/0 & 0/0 & 0/0 & 0/1 \\ 
netcounter & 2454 & 44 & 69 & 68 & 44 &\cellcolor{gray!30} 77  & 0/0 & 0/0 & 0/1 & 0/0 & 0/1 \\ 
alarmclock & 2491 & 66 & 67 & 68 & 57 &\cellcolor{gray!30} 71  &\cellcolor{gray!30} 1/1 & 0/0 &\cellcolor{gray!30} 1/3 &\cellcolor{gray!30} 1/1 & 0/2 \\ 
Nectroid & 2536 & 34 & 32 & 56 & 62 &\cellcolor{gray!30} 71  & 0/0 & 0/0 & 0/1 & 0/0 & 0/1 \\ 
QuickSettings & 2934 &\cellcolor{gray!30} 53 & 44 & 38 & 47 & 48 & 0/0 & 0/0 & 0/0 & 0/0 &\cellcolor{gray!30} 1/1 \\ 
MyExpenses & 2935 & 46 & 45 & 34 & 38 &\cellcolor{gray!30} 59  & 0/0 & 0/0 &\cellcolor{gray!30} 1/2 & 0/0 &\cellcolor{gray!30} 1/2 \\ 
a2dp & 3523 & 43 & 36 & 40 & 32 &\cellcolor{gray!30} 47  & 0/0 & 0/0 & 0/0 & 0/0 &\cellcolor{gray!30} 1/3 \\ 
mnv & 3673 & 18 & 28 &\cellcolor{gray!30} 48 & 12 & 43  &\cellcolor{gray!30} 2/2 & 1/1 & 0/1 & 0/0 & 1/2 \\ 
hotdeath & 3902 & 43 & 64 & 50 & 69 &\cellcolor{gray!30} 75  &\cellcolor{gray!30} 1/1 & 0/0 & 0/0 & 0/0 & \cellcolor{gray!30} 1/1 \\ 
SyncMyPix & 4104 & 21 & 20 & 25 & 21 &\cellcolor{gray!30} 26  & 0/0 & 0/0 & 0/1 & 0/0 & 0/1 \\ 
jamendo & 4430 & 21 & 23 & 16 & 24 &\cellcolor{gray!30} 28  & 0/0 & 0/0 &\cellcolor{gray!30} 1/2 & 0/0 & \cellcolor{gray!30} 1/3 \\ 
mileage & 4628 & $-$ & 28 & 26 & 35 &\cellcolor{gray!30} 62  & 1/1 & 1/1 & 1/3 & 1/1 &\cellcolor{gray!30} 2/4 \\ 
sanity & 4840 & 15 & 15 & 22 & 15 &\cellcolor{gray!30} 35  &\cellcolor{gray!30} 1/1 & 0/0 & 0/1 &\cellcolor{gray!30} 1/2 &\cellcolor{gray!30} 1/2 \\ 
fantastichmemo & 8419 & 40 & 29 & 29 & 25 &\cellcolor{gray!30} 41  & 0/0 &\cellcolor{gray!30} 1/1 & 0/1 & 0/0 &\cellcolor{gray!30} 1/3 \\ 
anymemo & 8428 & 25 & 30 & 32 & 20 &\cellcolor{gray!30} 43  & 1/1 & 0/0 & 1/2 & 1/1 &\cellcolor{gray!30} 3/4 \\ 
Book$-$Catalogue & 9857 & 34 & 9 & 14 & 22 &\cellcolor{gray!30} 35  &\cellcolor{gray!30} 2/2 & 1/1 & 0/1 & 1/1 &\cellcolor{gray!30} 2/3 \\ 
Wordpress & 10100 & 5 & 4 & 4 & 4 &\cellcolor{gray!30} 6 & 0/0 & 0/0 &\cellcolor{gray!30} 2/3 & 0/0 & 0/2 \\ 
passwordmanager & 10833  & 11 & 7 &\cellcolor{gray!30} 12 & 5 & 8  & 0/0 & 0/0 & 0/0 & 0/0 & 0/1 \\ 
aagtl & 11724 & 16 & 17 & 15 &\cellcolor{gray!30} 18 &\cellcolor{gray!30} 18  &\cellcolor{gray!30} 2/2 & 1/1 &\cellcolor{gray!30} 2/2 & 1/1 &\cellcolor{gray!30} 2/2 \\ 
morphoss & 17148 & 10 & 18 & 17 & 15 &\cellcolor{gray!30} 25  &\cellcolor{gray!30} 2/2 & 0/0 & 0/0 & 0/0 & 1/3 \\ 
addi & 19945 & 14 & 18 & 17 &\cellcolor{gray!30} 19 & 18 &\cellcolor{gray!30} 2/2 & 1/1 & 1/2 & 1/1 & 1/2 \\ 
k9mail & 22208 & 5 & 7 &\cellcolor{gray!30} 8 & 5 & 7  & 0/0 & 0/0 & 0/2 &\cellcolor{gray!30} 1/1 & 0/2 \\ 
\hline
Overall &  & 40.3 & 42.3 & 42.8 & 42.6 &\cellcolor{gray!30} 49 &24/25  &16/18  &26/62  &19/21  &\cellcolor{gray!30} 41/87
\\
\hline
\end{tabular}
}
\begin{flushleft}
DD.= \Name{}.
St.= Stoat
Mon.= Monkey
Sap.= Sapienz
``-"=Not applicable(crash emulator)
"a(b)" indicates a=\#crashes without system intent (b=\#crashes for all)
\end{flushleft}
\normalsize
\vspace*{-5pt}
\end{table}


}

\begin{table*}[t]
\centering
\scriptsize
\setlength{\tabcolsep}{2pt}
\caption{\label{tab:coverage} Testing Result for Comparison}
\scalebox{0.9}{
\begin{tabular}{|l|c|c c c c c c|c c c c c c|} \hline
  \#{\em APP.}
& {\em \# LOC}  
& \multicolumn{6}{c|}{\em Code Coverage}
& \multicolumn{6}{c|}{\em \# Fault Triggered}
\\
&
& {\em Mon.} &{\em Qt.} &{\em QBE.} &{\ \em St.\ } & {\em Sap.}  & {\em DD.}  
& {\em Mon.} &{\em Qt.} &{\em QBE.} &{\ \em St.\ } & {\em Sap.}  & {\em DD.}  
\\ \hline
soundboard & 109  & 31 & 42  & 42	& 42 & \cellcolor{gray!30}56 & 42 & 0/0 & 0/0 & 0/0 & 0/0 & 0/0 & 0/0 \\ 
gestures & 121 & 26 & 32 & 32 & 32 & \cellcolor{gray!30}51 & 32  & 0/0 & 0/0 & 0/0 & 0/0 & 0/0 & 0/0 \\ 
fileexplorer & 126  & 31 & \cellcolor{gray!30}40 & \cellcolor{gray!30}40 & \cellcolor{gray!30}40 & 31 & \cellcolor{gray!30}40 & 0/0 & 0/0 & 0/0 & 0/0 & 0/0 & 0/0  \\ 
adsdroid & 236 & 13 & \cellcolor{gray!30}40 & 29 & 29 & 33 &25  & 3/3 & 3/3 & 0/0 & \cellcolor{gray!30}5/2 &3/2 & 3/0 \\ 
MunchLife & 254 & 65 & 63 & 65 & 65 & \cellcolor{gray!30}71 & 66  & 0/0 & 0/0 & 0/0 & 0/0 & 0/0 & 0/0 \\ 
Amazed & 340  & 66 & 58 & 69 & 58 & 71 & \cellcolor{gray!30}73  & \cellcolor{gray!30}6/6 & 0/0 & 0/0 & 0/0 & 3/3 & 1/1 \\ 
battery & 342 & 72	& 71 & 69 & 71 & 54 & \cellcolor{gray!30}73  & 0/0 & 0/0 & 0/0 & 3/3 & 1/1 & \cellcolor{gray!30}9/9 \\ 
manpages & 385 & 58 & 58 & 62 & 48 & \cellcolor{gray!30}67 & 63  & 0/0 & 0/0 & 0/0 & \cellcolor{gray!30}3/0 & 0/0 & \cellcolor{gray!30}3/0 \\ 
RandomMusicPlayer & 400 & 52 & 70 & 65 & \cellcolor{gray!30}74 & 51	& 60  & 0/0 & 0/0 & 0/0 & \cellcolor{gray!30}3/0 & 0/0 & \cellcolor{gray!30}3/0 \\ 
AnyCut & 436 & 61 & 60 &62 & 61 & \cellcolor{gray!30}65	& 62  & 0/0 & 0/0 & 0/0 & 0/0 & 0/0 & 0/0 \\ 
autoanswer & 479 & 11 & 13 &13 & \cellcolor{gray!30}26 & 16 & 24 & 0/0 & 0/0 & 0/0 & 4/1 & 0/0 & \cellcolor{gray!30}5/2 \\ 
LNM & 492 & 54	& 58 & 56 & \cellcolor{gray!30}62 & 55 & 60  & 0/0 & 0/0 & 0/0 & \cellcolor{gray!30}6/0 & 1/1 & 3/0 \\ 
baterydog & 556 & 62 & 55 & 62 & 54 & \cellcolor{gray!30}67	& 62  & \cellcolor{gray!30}1/1 & 0/0 & 0/0 &  \cellcolor{gray!30}1/1 & 0/0 & 0/0 \\ 
yahtzee & 597 & 38 & 10 & 43	& 49 & 43 & \cellcolor{gray!30}57  & 1/1 & 0/0 & 3/3 & 3/0 & 0/0 & \cellcolor{gray!30}6/3\\ 
LolcatBuilder & 646 & 25 & 25 & 20 & 25 & 27	& \cellcolor{gray!30}52 & 0/0 & 0/0 & 0/0 & 0/0 & 0/0 & 0/0 \\ 
CounterdownTimer & 650 & 58 & 60 & 69 & \cellcolor{gray!30}77 & 45 & \cellcolor{gray!30}77  & 0/0 & 0/0 & 0/0 & 0/0 & 0/0 & 0/0 \\ 
lockpatterngenerator & 669  & 78 & 69 	& 75	& 66	& \cellcolor{gray!30}79	& 78  & 0/0 & 0/0 & 0/0 & 0/0 & 0/0 & 0/0 \\ 
whohasmystuff & 729 & 46 & 65 & 66 & 67 & 32	& \cellcolor{gray!30}74  & \cellcolor{gray!30}3/3 & 0/0 & 0/0 & \cellcolor{gray!30}3/0 & 0/0 & \cellcolor{gray!30}3/0\\ 
Translate & 799 & 45	& 44 & 43	& 40	& \cellcolor{gray!30}48	& 47  & 0/0 & 0/0 & 0/0 & 0/0 & 0/0 & 0/0 \\ 
wikipedia & 809 & 25	& 27 & 26	& 27	& \cellcolor{gray!30}29	& 27  & 0/0 & 0/0 & 0/0 & 0/0 & 0/0 & \cellcolor{gray!30}6/3 \\ 
DivideAndConquer & 814  & \cellcolor{gray!30}83	& 53 & 79	& 52	& 80	& 58  & 0/0 & 0/0 & 0/0 & 0/0 & \cellcolor{gray!30}3/3 & 0/0 \\ 
zooborns & 817 & 34 & 20 & 29 & 35 & \cellcolor{gray!30}36	& 35  & 0/0 & 3/3 & 0/0 & \cellcolor{gray!30}4/1 & 0/0 & 3/0 \\ 
multismssender & 828 & 49 & 29 & 48	& 49 & 60 & \cellcolor{gray!30}68  & 0/0 & 0/0 & 0/0 & \cellcolor{gray!30}3/0 & 0/0 & \cellcolor{gray!30}3/0 \\ 
Mirrored & 862 & 44 & \cellcolor{gray!30}45 & \cellcolor{gray!30}45 & \cellcolor{gray!30}45 &	\cellcolor{gray!30}45 & \cellcolor{gray!30}45  & 0/0 & 0/0 & 0/0 & \cellcolor{gray!30}3/0 & 0/0 & \cellcolor{gray!30}3/0 \\ 
myLock & 885  & 26	& 29 & 27 & \cellcolor{gray!30}44 & 30 & 42  & 0/0 & 0/0 & 0/0 & 4/1 & 0/0 & \cellcolor{gray!30}6/3 \\ 
aLogCat & 901 & 66 & 65 & 67 & 70 & 40 & \cellcolor{gray!30}79  & 0/0 & 0/0 & 0/0 & 0/0 & 0/0 & 0/0 \\ 
aGrep & 928 & 45 & 38 & 49 & 37 & 54	& \cellcolor{gray!30}57  & 1/0 & 3/3 & \cellcolor{gray!30}9/6 & 2/2 & 3/3 & \cellcolor{gray!30}9/6 \\ 
dialer2 & 978 & 37	& 36 & 38 & 34 & 35 & \cellcolor{gray!30}46  & 0/0 & 0/0 & 0/0 & \cellcolor{gray!30}3/0 & 1/1 & \cellcolor{gray!30}3/0 \\ 
hndroid & 1038 & 7 & 10 & \cellcolor{gray!30}11	& 9	& 8	& 9  & 3/3 & 3/3 & 2/2 & 3/3 & 3/3 & \cellcolor{gray!30}4/3 \\ 
Bites & 1060 & 32 & 36 & 28 & 42	& 39 & \cellcolor{gray!30}50  & 3/3 & 3/3 & 1/1 & 11/8 & 3/3 & \cellcolor{gray!30}14/11 \\ 
tippy & 1083 & 82 & 75 & 60 & 74	& 80 & \cellcolor{gray!30}83  & 0/0 & 0/0 & 0/0 & 0/0 & 0/0 & 0/0 \\ 
weight$-$chart & 1116 & 38 & 40 & 61 & 41 & 55 & \cellcolor{gray!30}75  & 2/2 & 2/2 & 1/1 & \cellcolor{gray!30}3/3 & 1/1 & \cellcolor{gray!30}3/3 \\ 
importcontacts & 1139 & 40 & 41 & \cellcolor{gray!30}63	& 37 & 42 & 41  & 0/0 & 0/0 & 0/0 & 0/0 & 0/0 & 0/0 \\ 
worldclock & 1242 & 89 & \cellcolor{gray!30}92 & \cellcolor{gray!30}92 & \cellcolor{gray!30}92 & 90 & 89 & 0/0 & 0/0 & 0/0 & \cellcolor{gray!30}3/0 & 0/0 & \cellcolor{gray!30}3/0 \\ 
blokish & 1245 & 41 & 36 & \cellcolor{gray!30}50 & 36 & 44 & \cellcolor{gray!30}50 & \cellcolor{gray!30}3/3 & 0/0 & \cellcolor{gray!30}3/3 & 0/0 & 1/1 & \cellcolor{gray!30}3/3 \\ 
aka & 1307 & 53 & 78 & \cellcolor{gray!30}80 & 61 & \cellcolor{gray!30}80 & 64  & 2/2 & 2/2 & 3/3 & 0/0 & 3/3 & \cellcolor{gray!30}9/9 \\ 
Photostream & 1375 & 20 & 22 & 14 & 24 & \cellcolor{gray!30}27 & 21  & 3/3 & 3/3 & 3/3 & \cellcolor{gray!30}9/6 & 3/3 & 6/3\\ 
dalvik$-$explorer & 1375 & 39 & 67 & 69 & 69 & \cellcolor{gray!30}70 & \cellcolor{gray!30}70  & 3/3 & 3/0 & 4/3 & 3/0 & 4/3 & \cellcolor{gray!30}6/3 \\ 
tomdroid & 1519 & 47 & \cellcolor{gray!30}54 & 50 & 52 & 50	& 53  & 0/0 & 0/0 & 0/0 & 1/1 & 0/0 & \cellcolor{gray!30}3/3 \\ 
PasswordMaker & 1535 & 57 & \cellcolor{gray!30}61 & 53	& 56 & 33 & 56  & 1/1 & 4/4 & 8/8 & \cellcolor{gray!30}12/9 & 3/3 & 11/8 \\ 
frozenbubble & 1706 & \cellcolor{gray!30}81 & 55 & 58 & 55 & 78 & 74  & 0/0 & 0/0 & 0/0 & 0/0 & 0/0 & 0/0 \\ 
aarddict & 2197 & 13 & \cellcolor{gray!30}31 & 28 & 30 & 14	& 18 & 0/0 & 0/0 & 0/0 & \cellcolor{gray!30}6/6 & 0/0 & 0/0 \\ 
swiftp & 2214  & \cellcolor{gray!30}13	& \cellcolor{gray!30}13 & 12 & \cellcolor{gray!30}13 & \cellcolor{gray!30}13 & \cellcolor{gray!30}13  & 0/0 & 0/0 & 0/0 & 0/0 & 0/0 & \cellcolor{gray!30}3/0 \\ 
netcounter & 2454 & 44 & 70 & 67 & 61 & 44 & \cellcolor{gray!30}76  & 0/0 & 1/0 & 0/0 & \cellcolor{gray!30}4/1 & 0/0 & 3/0 \\ 
alarmclock & 2491 & 64	& 67 & 64 & 67 & 37 & \cellcolor{gray!30}70  & 3/3 & 1/1 & 0/0 & \cellcolor{gray!30}8/2 & 3/3 & 7/1 \\ 
Nectroid & 2536 & 34 & 64 & 31 & 58 & 57 & \cellcolor{gray!30}70  & 1/1 & 1/1 & 0/0 & 3/0 & 1/1 & \cellcolor{gray!30}5/2 \\ 
QuickSettings & 2934 & \cellcolor{gray!30}52 & 38 & 41 & 38	& 48 & 48 & 0/0 & 0/0 & 0/0 & 0/0 & 0/0 & \cellcolor{gray!30}3/3 \\ 
MyExpenses & 2935 & 48 & 38 & 42	& 42 & 20 & \cellcolor{gray!30}61  & 0/0 & 0/0 & 0/0 & \cellcolor{gray!30}4/1 & 0/0 & 3/1 \\ 
a2dp & 3523 & 43 & 32 & 38 & 41 & 30	& \cellcolor{gray!30}45  & 0/0 & 0/0 & 0/0 & 1/0 & 0/0 & \cellcolor{gray!30}9/3 \\ 
mnv & 3673 & 22 & 44 & 34 & 43 & 8 & \cellcolor{gray!30}45  & 3/3 & 2/2 & 3/3 & 3/3 & 1/1 & \cellcolor{gray!30}6/3 \\ 
hotdeath & 3902 & 62 & 52 & 64 & 49 & 64 & \cellcolor{gray!30}75  & 1/1 & 1/1 & \cellcolor{gray!30}2/2 & 0/0 & 0/0 & \cellcolor{gray!30}2/2 \\ 
SyncMyPix & 4104 & 21 & 20 & 20 & 25 & 21 & \cellcolor{gray!30}26 & 0/0 & 0/0 & 0/0 & \cellcolor{gray!30}3/0 & 0/0 & \cellcolor{gray!30}3/0 \\ 
jamendo & 4430 & 21 & 16 & 24 & 14 & 23 & \cellcolor{gray!30}26  & 1/1 & 1/1 & 2/2 & 7/3 & 0/0 & \cellcolor{gray!30}9/3 \\ 
mileage & 4628 & 19 & 37 & 32 & 31 & 36 & \cellcolor{gray!30}60  & 4/4 & 1/1 & 1/1 & 10/4 & 3/3 & \cellcolor{gray!30}12/5 \\ 
sanity & 4840 & 19 & 25 & 16 & 22 & 15 & \cellcolor{gray!30}34  & 2/2 & 3/0 & 2/0 & 2/0 & 4/1 & \cellcolor{gray!30}5/2 \\ 
fantastichmemo & 8419 & 24 & 39 & 32 & 22 & 28 & \cellcolor{gray!30}43  & 3/3 & 4/4 & 1/1 & 5/2 & 1/1 & \cellcolor{gray!30}9/3 \\ 
anymemo & 8428 & 30 & 41 & 30 &30 & 27 & \cellcolor{gray!30}43  & 1/1 & 2/2 & 0/0 & 4/1 & 1/1 & \cellcolor{gray!30}10/6\\ 
Book$-$Catalogue & 9857 & 34 & \cellcolor{gray!30}37 & 25 & 11 & 24 & 32  & 3/3 & 1/1 & 4/3 & 3/0 & 1/1 & \cellcolor{gray!30}10/7 \\ 
Wordpress & 10100 & 5 & 3 & 4 & 5 & 4 & \cellcolor{gray!30}8 & 0/0 & 0/0 & 0/0 & 9/6 & 0/0 & \cellcolor{gray!30}10/4 \\ 
passwordmanager & 10833 & \cellcolor{gray!30}10 & 3 & 5 & 7 & 4 & 8  & 1/1 & 0/0 & 0/0 & 0/0 & 0/0 & \cellcolor{gray!30}2/0 \\ 
aagtl & 11724 & 15 & \cellcolor{gray!30}18  & 16 & 16 & 17 & 17  & 5/5 & 4/4 & 3/3 & 4/4 & \cellcolor{gray!30}6/6 & 4/4 \\ 
morphoss & 17148 & 12 & 17 & 19 & 18 & 11 & \cellcolor{gray!30}25  & 4/4 & 0/0 & 1/1 & 6/3 & 2/2 & \cellcolor{gray!30}10/4 \\ 
addi & 19945 & 14 & 16 & 18 & 17	& \cellcolor{gray!30}19 & 18 & \cellcolor{gray!30}6/6 & 3/3 & 3/3 & \cellcolor{gray!30}6/3 & 3/3 & \cellcolor{gray!30}6/3 \\ 
k9mail & 22208 & 5	& 7 & \cellcolor{gray!30}11 & 7 & 5	 & 7  & 0/0 & 0/0 & 0/0 & 6/1 & 1/1 & \cellcolor{gray!30}7/1 \\ 
\hline
Overall &  & 39.8 & 41.5 & 43 & 42.1 & 41.2 &\cellcolor{gray!30}48.9 &73/72  & 54/47 &59/52  &189/78  &63/59  &\cellcolor{gray!30}269/130
\\
\hline
\end{tabular}
}
\begin{flushleft}
DD.= \Name{}.
St.= Stoat
Mon.= Monkey
Sap.= Sapienz
Qt.=Q-testing
"a/b" indicates a=\#crashes for all and b=\#crashes without system level events
\end{flushleft}
\normalsize
\end{table*}

\ignore{
\begin{figure*}[t]
\centering
  
  \begin{subfigure}[t]{0.14\textwidth}\hfill
    \includegraphics[scale=0.14]{figures/a1}
    \caption{}
    \label{subfig:a}
  \end{subfigure}
\vspace*{5pt}
\hspace*{10pt}
     \begin{subfigure}[t]{0.14\textwidth}\hfill
    	\includegraphics[scale=0.14]{figures/a2}
	\caption{}
	\label{subfig:b}
  	\end{subfigure}
\vspace*{5pt}
\hspace*{10pt}
     \begin{subfigure}[t]{0.14\textwidth}\hfill
    	\includegraphics[scale=0.14]{figures/a3}
	\caption{}
	\label{subfig:c}
  	\end{subfigure}	
\hspace*{10pt}
     \begin{subfigure}[t]{0.14\textwidth}\hfill
    	\includegraphics[scale=0.14]{figures/a4}
	\caption{}
	\label{subfig:d}
  	\end{subfigure}	
\hspace*{10pt}
     \begin{subfigure}[t]{0.14\textwidth}\hfill
    	\includegraphics[scale=0.14]{figures/a5}
	\caption{}
	\label{subfig:e}
  	\end{subfigure}	
\hspace*{10pt}
     \begin{subfigure}[t]{0.14\textwidth}\hfill
    	\includegraphics[scale=0.14]{figures/a6}
	\caption{}
	\label{subfig:f}
  	\end{subfigure}
	
   
 	 \vspace*{-10pt}
 \caption{\small \textbf{The steps of reproducing the 
 crash described in Fig.~\ref{fig:bugreport}.}}
	\label{fig:steps}
	\vspace*{-15pt}
\end{figure*}

}

\begin{figure*}[t]
    \hspace*{-15pt}
  \begin{minipage}[t]{0.20\textwidth}
    \label{fig:I2_coverage}
    \includegraphics[scale=0.3]{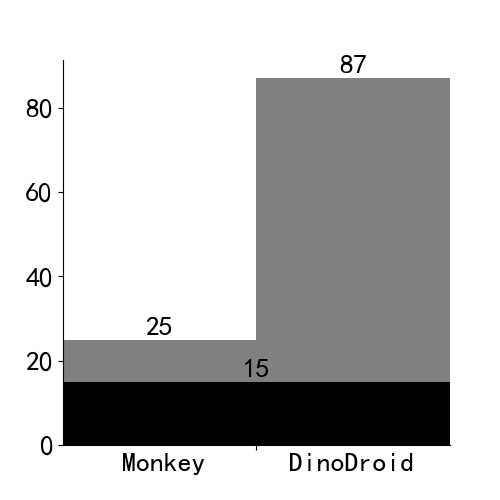}
  \end{minipage}%
    \begin{minipage}[t]{0.20\textwidth}
    \label{fig:Qtest_coverage}
    \includegraphics[scale=0.3]{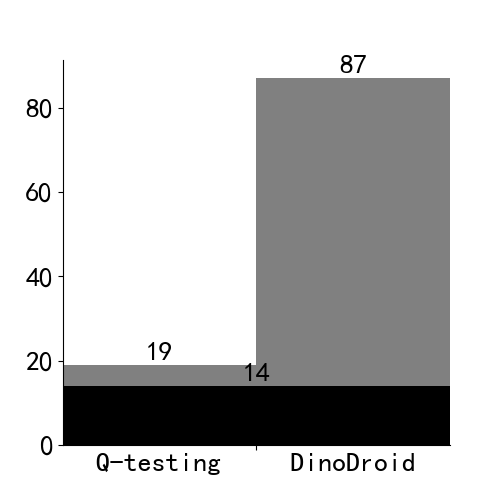}
  \end{minipage}%
  \begin{minipage}[t]{0.20\textwidth}
    \label{fig:St_coverage}
    \includegraphics[scale=0.3]{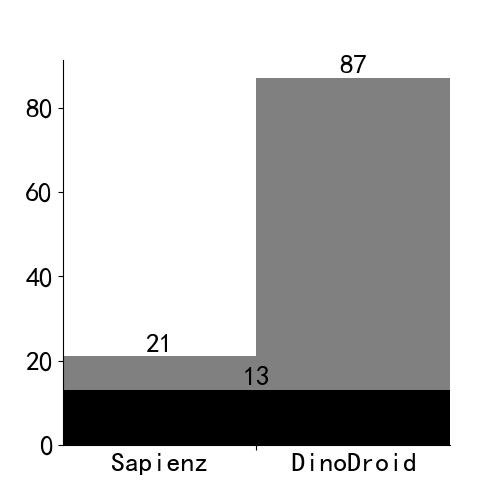}
  \end{minipage}
  \begin{minipage}[t]{0.20\textwidth}
    \label{fig:I2_coverage}
    \includegraphics[scale=0.3]{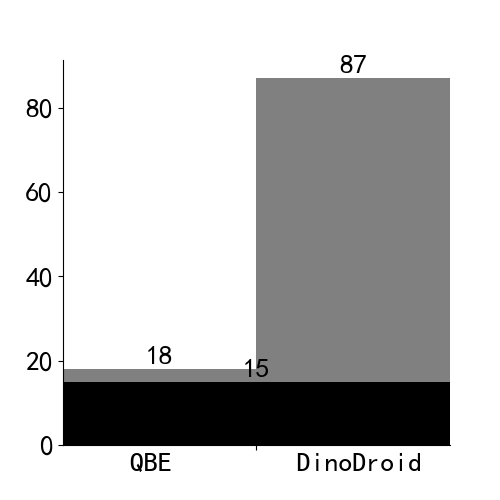}
  \end{minipage}%
  \begin{minipage}[t]{0.20\textwidth}
    \label{fig:St_coverage}
    \includegraphics[scale=0.3]{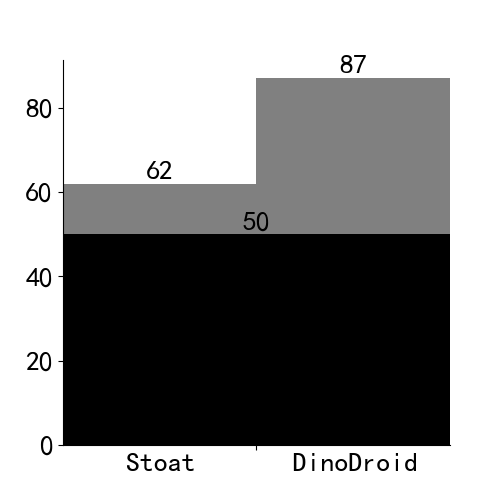}
  \end{minipage}
\caption{\textbf{Comparison of tools in detecting crashes}}
\label{fig:crash}
\end{figure*}

\ignore{
\begin{figure}[t]
  \centering
  \begin{minipage}[c]{0.25\textwidth}
    \label{fig:I2_coverage}
    \centering
    \includegraphics[scale=0.3]{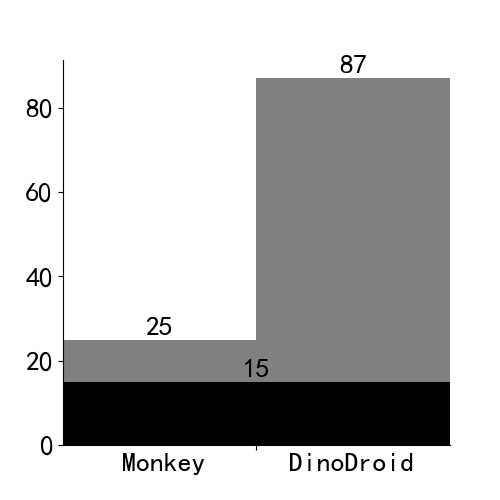}
  \end{minipage}%
  \begin{minipage}[c]{0.25\textwidth}
    \label{fig:St_coverage}
    \centering
    \includegraphics[scale=0.3]{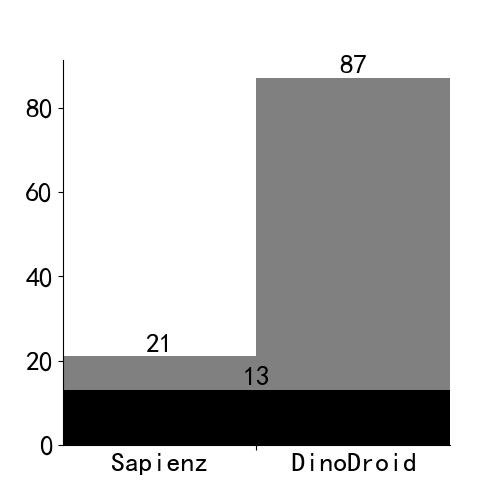}
  \end{minipage}

  \begin{minipage}[c]{0.25\textwidth}
    \label{fig:I2_coverage}
    \centering
    \includegraphics[scale=0.3]{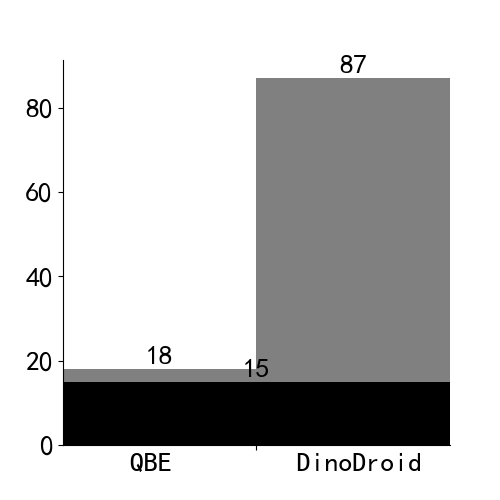}
  \end{minipage}%
  \begin{minipage}[c]{0.25\textwidth}
    \label{fig:St_coverage}
    \centering
    \includegraphics[scale=0.3]{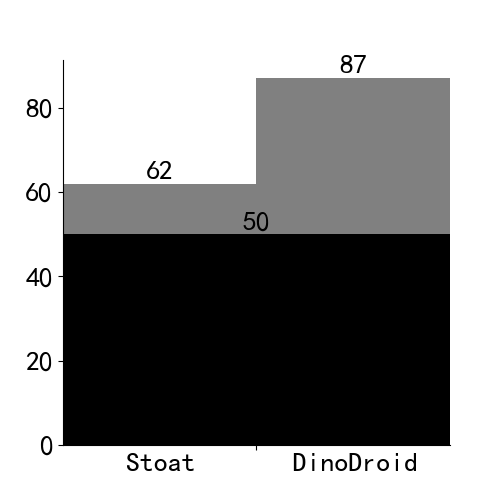}
  \end{minipage}

\caption{\textbf{Comparison of tools in detecting crashes}}
\label{fig:crash}
\end{figure}

}

\ignore{
\begin{figure}[t]
  \centering
  \hspace{-15pt}
  \begin{minipage}[c]{0.25\textwidth}
    \label{fig:I2_coverage}
    \centering
    \includegraphics[scale=0.3]{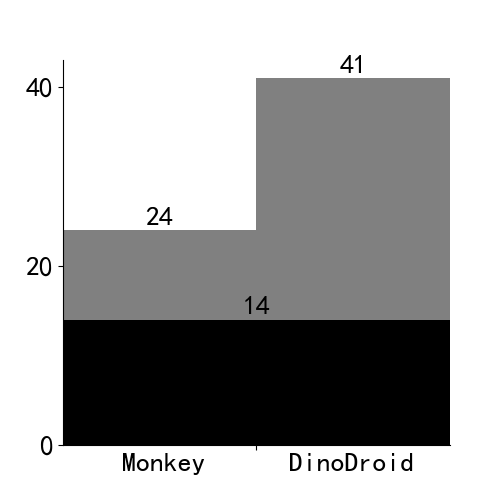}
  \end{minipage}%
  \begin{minipage}[c]{0.25\textwidth}
    \label{fig:St_coverage}
    \centering
    \includegraphics[scale=0.3]{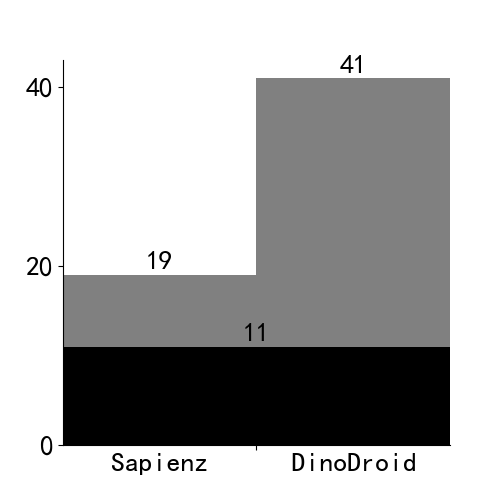}
  \end{minipage}

  \begin{minipage}[c]{0.25\textwidth}
    \label{fig:I2_coverage}
    \centering
    \includegraphics[scale=0.3]{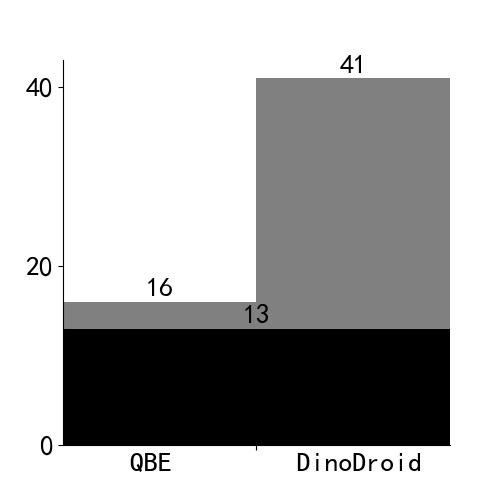}
  \end{minipage}%
  \begin{minipage}[c]{0.25\textwidth}
    \label{fig:St_coverage}
    \centering
    \includegraphics[scale=0.3]{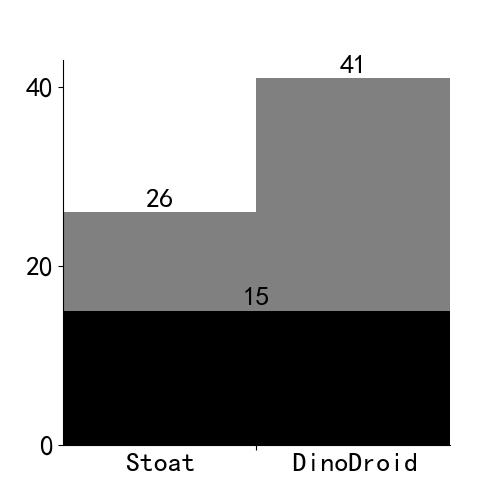}
  \end{minipage}

\caption{\textbf{Comparison of tools in detecting crashes filter out activity intent}}
\label{fig:crash1}
\end{figure}
}

\subsection{RQ3: Understanding the Learned Model}

RQ3 is used to understand whether the DQN agent 
can correctly learn the app behaviors 
based on the provided features. 
To do this, we analyzed the logs generated by the DQN agent 
for the 64 apps. 


\subsubsection{Understanding the Behaviors 
of Individual Features}

\subB{FCR feature}. We  first would like to understand the behavior
of the FCR feature. Specifically, we would like to know
whether \Name{}
will assign a higher priority to unexecuted events over executed events in the current page. 
We hereby computed the percentage
of pages performing expected actions (i.e., triggering unexecuted events) among 
all pages. 
We consider only unexecuted and executed events  instead of execution frequency
because we need to silent the influence of children pages (i.e., the FCD feature) as unexecuted
events do not invoke children pages. 
In total, there are 16,169 pages
containing both executed and unexecuted events, with a total of
197,279 events; 85.2\% of the pages performed expected
actions. 
We then computed the probability of randomly selecting an event from each page and found that only 51.6\% of the pages performed expected actions.
The results show that \Name{} indeed behaves better than a random approach. 

\ignore{
Note that it is possible that the executed  and unexecuted events 
are imbalance, e.g., 
there are significantly more unexecuted events on 
every page, so the results could be biased because the possibility
of selecting unexecuted events is higher. We randomly 
selected an event from each page and found that 51.6\% (8352/16,169) of the pages
performed expected actions.  This shows that our data is balanced and
\Name{} indeed behaves better than a random selection
approach. }

\subB{FCD feature}. 
We  next would like to understand the behavior
of the FCD feature.  Specifically, we would like to 
know whether \Name{} tends to 
select events with unexecuted children events
over those with executed children events. 
To control the variation of event execution frequencies (FCR), 
we selected only pages that contain events with the same execution 
frequency from the traces. As a result, 495 pages with a total of 3,182
events were selected and 81.2 \%  of the pages performed 
expected actions (i.e., selecting events
with unexecuted children events).
When using a random selection,  only 33.6\% pages
performed expected actions.


\ignore{For the question 3, to reduce the children feature noise, we change this question to "for the pages in which all events are executed and executed freq are same, whether DinoDroid executes event which has un-executed child". After the changing, we remove the affection of executed freq feature, because we only select pages which have same executed freq events. We select 494 pages which satisfy the changing requirement.  As shown in the Tab ~\ref{tab:human1}, in 81.2 \% pages DinoDroid does a right behavior. Random method only achieves 33.6 \%, because less events have unexecuted child in the data set than the events have executed child.
\commentty{pages containing widgets with the same times. first generation: correct if no executed children.}
}

\subB{TXC feature}. 
We also examined if \Name{} is able to understand
the content of events. To do this,  we control
the variations of FCR and FCD features
by selecting pages that do not contain any 
executed events from the traces. A 
 total of 3,060 pages were selected. We hypothesize that if Q values
of these events are different (i.e.,
their textual content is different), \Name{} is able
to recognize the content of the widgets 
(i.e., perform expect actions).
The results show that  91\% of pages performed
expected actions with the difference between highest Q value and lowest Q value greater than 10\%.


The above results suggest that \emph{the behaviors of the app features
learned by \Name{} are mostly expected.}

\subsubsection{Understanding the Effectiveness
of Individual Features}

\yuc{
We next explore how each of the three features
affects the coverage effectiveness of 
\Name{}. To achieve this, we repeated the
experiment but disabled the  
FCR feature, FCD feature, and TXC feature separately.
The results suggest that \Name{}'s average  code coverage is higher than the coverage of the absence of FCR, FCD, and TXC by 0.7\%, 2.4\%, 3\%, respectively.}
%
%

\yuc{We found the results surprising.  Prior to the experiment, we expected the FCR feature to be the most crucial feature during testing because using event frequency to guide the exploration is intuitive and has been a common
strategy used in existing work~\cite{su2017guided}. 
However, the actual result suggests that
the textual feature (TXT) contributes more than the other features. 
Therefore, it is very important for a tester or a tool to understand the meaning of text. }

\balance
\subsubsection{The Whole DQN Model Behaviors}

The above experiment suggests how individual 
features affect the learning process of \Name{}. 
We next conducted a deeper analysis to understand
the behaviors of \Name{} under the combination of the three features. 
To do this, we randomly selected a model from the two models
learned from the 64 apps (by two cross-fold validation).  
We used this model to process the first page of the 32 apps, which were not used to train the model.
We then manually set the values of FCR and TXC features while
keeping the TXC feature as default and see how 
the model behaves. 

\ignore{All of the experiments are in partial data sets in order to reduce the noise from unrelated features. In order to understand DinoDroid's behavior in comprehensive, we only focus on analysis pairs of features to reduce the noise from other features. Additionally, to analysis the behavior of DinoDroid, we focus on events with different features on one fixed model. This setting can eliminate the noise from multiple models in the real testing.  We randomly pick one from two generated models in above coverage testing as the fixed model. We select 34 launching pages from 34 apps which are from the testing set of the selected model in above coverage testing. 
In these 34 launching pages, there are 283 events. Every event keeps its own text feature. The executed freq and descendant features for each events are manually set. DinoDroid can provide a priority value(Q value) for each feature combination. DinoDroid always compute these higher priority values and pick the largest one in a page to execute in the testing. We take average of priority value from 283 events on every feature combination. Based on the average priority value, we can analysis the behavior of DinoDroid by different combination of features. There are some interesting behaviors never be revealed or designed for a heuristic by any existing works.
}

As shown in Table~\ref{tab:human2}, ``FCR" indicates the execution frequency of the 
current event is executed and ``FCD" indicates the three generations of children pages with
the number of unexecuted events and the number of events being executed once (i.e., the first two elements in the feature vector).  
Each number in the table is the Q value averaged across all events with the same feature setting. 
For example,  $<$\textit(6\#1); (1\#1); (1\#1)$>$ +  ``1" (the second row + the third column) indicates that
when an event is executed once, the first generation of its children page contains six unexecuted events and one event is executed once, 
and the second and third generation of
its children pages have one unexecuted events and one event is executed once.

\Name{} learned the following behaviors as shown in Table~\ref{tab:human2}. 
First,  the Q value of (6\#1);(1\#1);(1\#1) is larger than that of (1\#6);(1\#1);(1\#1). 
This indicates that an event with more unexecuted children events in the first
generation is more likely to be selected. Second, the Q value of (1\#1);(6\#1);(1\#1)
is larger than that of (1\#1);(1\#6);(1\#1) and the Q value of (1\#1); (1\#1); (6\#1)
is larger than that of (1\#1); (1\#1); (1\#6). This indicates the second and third generation 
(or perhaps the subsequence generations) of children events can also guide
the exploration like the first generation. Third, the unexecuted event on the current page has the highest priority to be selected since its Q value is significantly larger (i.e., FCR=0). 
Fourth, with the same FCD feature values, the event with less execution frequency on the current page has a higher priority to be selected since the Q value is decreasing as the value of FCR increases.

\ignore{
As shown in Table~\ref{tab:human2}, ``Times" indicates the number of times the current widget is executed and ``Children" indicates the three generations of children pages with
the number of unexecuted widgets and the number of widgets being executed once (i.e., the first two elements in the feature vector).  
Each number in the table is the Q value averaged across all events with the same feature setting. 
setting. For example,  Row 11\textit{(5\#1); (0\#0); (0\#0)}" + Column ``1" indicates that
when a widget is executed one time, the first generation of its children page contains five unexecuted widgets and one widget is executed once, 
and the second and third generation of
its children pages do not contain  any unexecuted widgets or widgets being executed 

\Name{} learned the following behaviors as shown in Table~\ref{tab:human2}. 
First,  the Q value of (5\#1);(0\#0);(0\#0) is larger than that of (1\#5);(0\#0);(0\#0). 
This indicates that a widget with more unexecuted children widgets in the first
generation is more likely to be selected. Second, the Q value of (0\#0);(5\#1);(0\#0)
is larger than that of (0\#0);(1\#5);(0\#0) and the Q value of (0\#0); (0\#0); (5\#1)
is larger than that of (0\#0); (0\#0); (1\#5). This indicates the second and third generation 
(or perhaps the subsequence generations) of children widgets can also guide
the exploration like the first generation. Third, the effects are decreasing
for later generations. For example the Q value of (5\#1);(0\#0);(0\#0) is 
larger than that of (0\#0);(5\#1);(0\#0), which is larger than (0\#0); (0\#0); (5\#1). 
Third, the unexecuted widget on the current page has the highest priority 
to be executed since its Q value is significantly larger (i.e., (0\#0);(0\#0);(0\#0)). 
Fourth, with the same FCD feature values, the widget with less execution frequency
in the current page has a higher priority to be executed since the Q value is 
decreasing as the value of FCR increases. 
}
All of the above behaviors  align with the intuition on \emph{how human
would test apps} with the three types of features and demonstrate that \Name{}'s DQN 
can \emph{automatically learn these behaviors without the need
to manually set  heuristic rules. 
}





\ignore{
1. On every executed freq (5\#1);(0\#0);(0\#0)'s priority value is higher than (1\#5);(0\#0);(0\#0). It means DinoDroid considers the higher number of unexecuted events in first generation will produce higher probability for the code coverage. This makes sense because more unexecuted events in first generation may result more potential functionality for current event.

2. The first result is not only work on the first generation. It also works on the second generation and third generation. For example, on every executed freq, (0\#0);(5\#1);(0\#0) has priority value than (0\#0);(1\#5);(0\#0). This makes sense because more unexecuted events in any generation may result more potential functionality for current event.

3. There is a decreasing affection of unexecuted events in deep generations. We can see (5\#1);(0\#0);(0\#0)'s priority value is higher than (0\#0);(5\#1);(0\#0) on every executed freq. It makes sense because the first generation is more near to the current page. The potential functionality increasing in first generation will be more direct to get. However, sometimes if the second generation has very big number of unexecuted events than first generation, it may have higher priority value than first generation. For example, (0\#0);(5\#1);(0\#0) has higher priority value than first generation.

4. Whatever the execution frequency and descentant feature, the unexecuted event on the current page has extreme high priority value. In the table, freq=0 and descentant=(0\#0);(0\#0);(0\#0) has priority value as 11.08 which is double for the maximum value from other combinations.

5. On the same descentant feature, the higher executed time will have small priority value. For example, (5\#1);(0\#0);(0\#0) on freq=1 is 4.48 and on freq=2 is 2.01. It because DinoDroid learns high executed frequency event may not produce new line coverage. The freq affects more on small executed freq than larget executed freq. For example, the difference between  freq=1 and freq=2 on (1\#5);(0\#0);(0\#0) is 3.69. The difference between freq=2 and freq=3 on same descentant feature is 0.6. DinoDroid may consider the high executed freq's result are not distinguished. For example, freq=3 and freq=4 are-0.59 and -0.56. 

6. The event without no any descentants have the lowest executed time in every freq.
}

\ignore{
\begin{table}[]
\caption{\label{tab:human1} Real trace evaluation for human behavior comparison}
\begin{tabular}{|l|l|l|l|}
\hline
Questions                                                                                                                      & DataSet                                                                              & Technology & Right  \\ \hline
\multirow{2}{*}{\begin{tabular}[c]{@{}l@{}}Question1: First to execute no \\executed event \end{tabular}}                & \multirow{2}{*}{\begin{tabular}[c]{@{}l@{}}16169 pages\\ 197279 events\end{tabular}} & DinoDroid  & 85.2\% \\ \cline{3-4} 
                                                                                                                               &                                                                                      & Random     & 52.9\% \\ \hline
\multirow{2}{*}{\begin{tabular}[c]{@{}l@{}}Question 2: Different priorities on \\different events \end{tabular}}                      & \multirow{2}{*}{\begin{tabular}[c]{@{}l@{}}3060 pages\\ 24999 events\end{tabular}}   & DinoDroid  & 98\%   \\ \cline{3-4} 
                                                                                                                               &                                                                                      & Random     & -      \\ \hline
\multirow{2}{*}{\begin{tabular}[c]{@{}l@{}}Question 3: High priority to \\ execute event with no executed child.\end{tabular}} & \multirow{2}{*}{\begin{tabular}[c]{@{}l@{}}495 pages\\ 3182 events\end{tabular}}     & DinoDroid  & 81.2\% \\ \cline{3-4} 
                                                                                                                               &                                                                                      & Random     & 33.6\% \\ \hline
\end{tabular}
\end{table}
}

\begin{table}[]
\caption{\label{tab:human2} DinoDroid's behavior on every feature combination}
\scalebox{0.9}{
\begin{tabular}{|l|c|c|c|c|c|c|}
\hline
\diagbox [width=8em,trim=l] {FCD}{FCR}                  & 0 & 1 & 2 & 3 & 4 &5\\ \hline
$<$(6\#1), (1\#1), (1\#1)$>$ & -   & 3.45  & 2.32  & 1.2 & -0.65 &-0.47  \\ \hline
$<$(1\#6); (1\#1); (1\#1)$>$ & -  &  2.57 & 0.23  & -0.171  & -3.35  &-3.55\\ \hline
$<$(1\#1); (6\#1); (1\#1)$>$ & -  & 3.44  &  2.03 & 0.16  & 0.19 & 0.00  \\ \hline
$<$(1\#1); (1\#6); (1\#1)$>$ & -  & 1.01 & 0.92  & -0.31  & -3.59 &-3.96  \\ \hline
$<$(1\#1); (1\#1); (6\#1)$>$ & -  & 1.92 & 0.97  & 1.07 & -0.9 &-3.2 \\ \hline
$<$(1\#1); (1\#1); (1\#6)$>$ & -  & 1.09 & 0.81 & 0.21  & -3.3 &-4.53 \\ \hline
$<$(1\#1); (1\#1); (1\#1)$>$ & -  & 1.01 & 0.78  & -1.66  & -3.86 &-4.32  \\ \hline
$<$(0\#0); (0\#0); (0\#0)$>$ & 10.96  & -4.42  & -4.98  & -5.09  & -5.13 &-5.14 \\ \hline
\end{tabular}
}
\begin{flushleft}
\small{$<$(A\#B); (C\#D); (E\#F)$>$: the first generation has $A$ unexecuted
events and $B$ events are executed once; the  second generation has $C$ unexecuted
events and $D$ events are executed once; the  third generation has $E$ unexecuted
events and $F$ events are executed once.}
\end{flushleft}
\normalsize
\end{table}


\section{Limitations and Threats to Validity}
\label{sec:discussion}

\Name{} has several limitations. First, \Name{} currently considers  three features. 
As part of the future work, we will assess whether other 
features, \yuc{such as images of widgets and
the execution of sequences of actions that compose complete use cases, can improve the performance of \Name{}.}
Second,  \Name{} handles feature values represented by
matrices or vectors. However,
some features may need more
complex representations (e.g., multiple complex 
matrices) and processing them may take substantial amount of time.
Third, in the current version, 
\Name{} measures line coverage to train the apps, 
so we focus on open-source apps.  
\Name{} can be extended to handle closed-source apps 
by collecting method-level coverage
or using non-code-related rewards.
\yuc{Fourth, 
for a fair comparison, 
\Name{} does not 
utilize extra advantageous features
(e.g. login script, short message records, contact records)
in its current version, because none of the 
state-of-the-art tools have
such features by default. It is also very challenging
to modify the tools to implement these features
because 
we cannot control their installation of apps and 
do not know when to run the login scripts.  
Nevertheless, we believe
that adding the extra functionality might increase the effectiveness
of coverage and bug detection in DinoDroid and  the existing tools. 
For example, after we added the login function in \Name{}, 
k9 mail's code coverage was increased to 40\% in one hour,
compared with 7\% coverage without the login function.
Su et al.~\cite{su2021benchmarking} implemented the login function and resolved 
some usability issues for the six tools in their study, 
but the apps in our benchmarks are emma instrumented and thus incompatible with their tools, which are 
jacoco instrumented. 
As part of future work, we will implement the 
extra advantageous features in DinoDroid, as well as in the state-of-the-art tools
in our study. We will also compare DinoDroid with more existing tools, such as 
those studied in~\cite{su2021benchmarking}. 
Fifth, the current version of our approach uses a fixed reward. 
Existing work (Q-testing~\cite{pan2020reinforcement}) has designed an adaptive reward function to exploit a neural network to calculated the difference between two states (app pages) to determine the reward value.
We believe that our
approach can work with the adaptive reward function by simply using Q-testing's trained model to determine the reward value. 
}

The primary threat to the external validity of this work involves
the representativeness of the apps used in our study. Other apps 
may exhibit different behaviors. We reduce this threat to some 
extent by using a set of apps developed and released by 
prior research work~\cite{choudhary2015automated}, which has been extensively
used by existing Android testing research~\cite{moran2016automatically,li2019humanoid}. 
These apps were also used by the baseline tools (i.e., Stoat~\cite{su2017guided}, Sapienz~\cite{mao2016sapienz}) in our study.

The primary threat to internal validity involves potential errors 
in the implementations of  \Name{}. We control this threat 
by testing our tools extensively and verifying their 
results against a small-scale program for which we
can manually determine the correct results.

\section{Related Work}
\label{sec:related}

There have been a number of techniques on automated Android GUI testing.
Random testing~\cite{monkey} uses 
random strategies to generate events.
Because of its simplicity and availability, it can send thousands of events per 
second to the apps and can thus get high code coverage. 
However, the generated events may be largely redundant 
and ineffective.  DynoDroid~\cite{machiry2013dynodroid} improved random testing
by exploring the app in a manner that can avoid testing redundant widgets.
But it may still be ineffective in reaching functionalities involving deep levels of the app
due to randomness. 
%
%
Sapienz~\cite{mao2016sapienz} uses multi-objective 
search-based testing to maximize coverage and fault revelation at the same time to 
optimize test sequences and minimize length. TimeMachine~\cite{dong2020time} enhances random testing with the ability to jump to a state in the past.

Model-based~\cite{amalfitano2012using, amalfitano2015mobiguitar, azim2013targeted, baek2016automated, gu2017aimdroid, yan2018land, yan2017widget, yang2013grey} build and use a GUI model of the app to generate tests. 
The models are usually represented by finite state machines to
model app states and their transitions. 
For example, Stoat ~\cite{su2017guided} utilizes a stochastic Finite State Machine model to describe the behavior of app under test.
Section~\ref{sec:introduction} discusses Stoat. 
Unlike model-based testing, \Name{} does not
need to model the app behaviors. Instead, it can automatically 
learn app behaviors by deep Q-networks. 

Systematic testing tools, such as symbolic execution~\cite{mahmood2014evodroid, gao2018android},
aims to generate test cases to cover some code 
that is hard to reach. While symbolic execution
may be able to exercise  functionalities that 
are hard to reach, it is less scalable
and not effective in code coverage or bug detection.

Machine learning techniques have also been used in testing
Android apps. These  techniques can be classified into two 
categories~\cite{pan2020reinforcement}. 
The techniques in the first category typically have an 
explicit training process to learn knowledge from existing apps 
then apply the learned experience on new 
apps~\cite{koroglu2018qbe, borges2018guiding, li2019humanoid, lin2019test, degott2019learning}. 
For example, QBE~\cite{koroglu2018qbe} learns how to test android apps in a training set 
by a Q-learning. As discussed in Section~\ref{subsec:limitations},
QBE is not able to capture fine-grained app behaviors due to the limitations
of Q-learning.  

In addition to QBE, Degott et al.~\cite{borges2018guiding} use learning to identify valid 
interactions for a GUI widget (e.g., whether a widget accepts interactions). It then uses
this information to guide app exploration. Their following work~\cite{degott2019learning} 
uses MBA reinforcement learning to guide the exploration based on
the abstracted features: valid interactions and invalid interactions. However, like QBE,
these techniques can not handle complex app features because of 
the high-level abstraction of state information. 

Humanoid ~\cite{li2019humanoid} employs deep learning to train a model
from labeled human-generated interaction traces  and uses the model together
with a set of heuristic rules to guide the exploration of new apps.
The result of deep learning is only used to select one event 
from the unexplored events on the current page. 
If all events on the current page are explored, 
the heuristic rules will be used to compute the shortest path in a graph to find 
a page with unexplored events.
In contrast, \Name{} does not need to manually build/label 
the training set or use any pre-defined search strategies or rules to guide testing.  
AppFlow ~\cite{hu2018appflow} generates
GUI tests from high-level, manually written test cases. 
It leverages machine learning
to match screens and widgets to individual tests.
Juha Eskonen et al.~\cite{eskonen2020automating}
employs deep reinforcement learning 
to perform UI testing for web applications.
This work specifically focuses on the image feature
and does not perform an an empirical study to assess
its performance.
In contrast, \Name{}'s feature merging component
is able to handle multiple complex features.
\Name{} targets at Android apps and performs
an extensive empirical study.


The techniques in the second category do not have explicit training 
process~\cite{adamo2018reinforcement, vuong2019semantic, vuong2018reinforcement, 
mariani2012autoblacktest, pan2020reinforcement, koroglu2021functional}.
Instead, they use Q-learning to guide the exploration of individual apps, where each app generates a unique behavior model. 
These techniques share the same limitations with
Q-learning, where complex features cannot be maintained
in the Q-table.  For example, QDROID~\cite{vuong2019semantic},  DRIFT~\cite{harries2020drift}, and ARES~\cite{romdhana2021deep} design the deep neural network agents to guide the exploration of Android apps. 
However, like QBE, the widgets are abstracted into several categories or boolean values. 
The abstraction of the states and actions may cause the loss of information in individual widgets. 
%
Wuji ~\cite{zheng2019wuji} combines reinforcement learning 
with evolutionary algorithms to test Android game apps. 
It designs a unique state vector for 
each game used as the position of the player 
character or the health points. Therefore, Wuji can not transfer
the learned knowledge to new apps. 
Q-testing~\cite{pan2020reinforcement}'s main contribution is
the design of a reward function to improve the performance of Q-testing
by calculating the difference between the current state and the recorded states.
Our work is orthogonal to Q-testing and one can improve the performance
of \Name{} by utilizing Q-testing's reward function.

\section{Conclusions}
\label{sec:conclusion}

We have presented \Name{}, an automated approach to test Android application. 
It is a  deep Q-learning-based approach, which can learn how to test android application 
rather than depending on heuristic rules. \Name{} can take more complex features 
than existing learning-based methods as the input by using a Deep Q learning-based structure. 
With these features, \Name{} can process complex features on the pages of an app.
Based on these complex features, \Name{} can learn a policy targeting at achieving high code coverage. 
We have evaluated \Name{} on 64 apps from a widely used benchmark and showed that \Name{} 
outperforms the state-of-the-art and state-of-practice Android GUI testing tools
 in both code coverage and bug detection. By analyzing the testing traces
 of the 64 apps, we are also able to  tell that the machine really
  understands the  features and provides a sensible strategy to 
  generate tests.

\ifCLASSOPTIONcompsoc
  \section*{Acknowledgments}
\else
  \section*{Acknowledgment}
\fi

This research is supported in part by the NSF grant CCF-1652149.

\ifCLASSOPTIONcaptionsoff
  \newpage
\fi

\bibliographystyle{IEEEtran}
\bibliography{IEEEabrv,bib/paper}
\end{document}